\documentclass[fleqn,usenatbib]{mnras}

\usepackage{newtxtext,newtxmath}

\usepackage[T1]{fontenc}
\usepackage{ae,aecompl}
\usepackage{tikz}
\usepackage{bm}
\usepackage{relsize}
\usetikzlibrary{positioning}
\usepackage{listings}
\usepackage{xcolor} 

\usepackage{graphicx}	
\usepackage{amsmath}	
\usepackage{amssymb}	
\usepackage{color}
\usepackage{cprotect}

\definecolor{linkcolor}{rgb}{0,0,0.25}
\hypersetup{
  colorlinks=true,        
  linkcolor=linkcolor,    
  citecolor=linkcolor,    
  filecolor=linkcolor,    
  urlcolor=linkcolor      
}

\makeatletter
\renewcommand{\@printed}{}
\makeatother

\definecolor{darkgreen}{rgb}{0.0, 0.5, 0.0}

\definecolor{darkblue}{rgb}{0.0, 0., 0.7}

\renewcommand{\vec}[1]{\ensuremath{\mathbf{#1}}}
\newcommand{\teff}{\ensuremath{T_\mathrm{eff}}}
\newcommand{\logg}{\ensuremath{\log g}}
\newcommand{\xh}[1]{\ensuremath{[\mathrm{#1/H}]}}
\newcommand{\xfe}[1]{\ensuremath{[\mathrm{#1/Fe}]}}
\newcommand{\madstd}{\ensuremath{\sigma^{\mathrm{MAD}}}}

\title[Deep learning of stellar abundances]{Deep learning of multi-element abundances from high-resolution spectroscopic data}

\author[Leung \& Bovy]{
Henry W. Leung$^{1}$\thanks{E-mail: henrysky.leung@mail.utoronto.ca} \&
Jo Bovy$^{1,2}$\thanks{Alfred P. Sloan Fellow}
\\
$^{1}$Department of Astronomy and Astrophysics, University of Toronto, 50 St. George Street, Toronto, Ontario, M5S 3H4, Canada\\
$^{2}$Dunlap Institute for Astronomy and Astrophysics, University of Toronto, 50 St. George Street, Toronto, Ontario, M5S 3H4, Canada
}

\date{}

\pubyear{2018}

\begin{document}      
\label{firstpage}
\pagerange{\pageref{firstpage}--\pageref{lastpage}}
\maketitle
\begin{abstract}
Deep learning with artificial neural networks is increasingly gaining attention, because of its potential for data-driven astronomy. However, this methodology usually does not provide uncertainties and does not deal with incompleteness and noise in the training data. In this work, we design a neural network for high-resolution spectroscopic analysis using APOGEE data that mimics the methodology of standard spectroscopic analyses: stellar parameters are determined using the full wavelength range, but individual element abundances use censored portions of the spectrum. We train this network with a customized objective function that deals with incomplete and noisy training data and apply dropout variational inference to derive uncertainties on our predictions. We determine parameters and abundances for 18 individual elements at the $\approx 0.03\,\mathrm{dex}$ level, even at low signal-to-noise ratio. We demonstrate that the uncertainties returned by our method are a realistic estimate of the precision and they automatically blow up when inputs or outputs outside of the training set are encountered, thus shielding users from unwanted extrapolation. By using standard deep-learning tools for GPU acceleration, our method is extremely fast, allowing analysis of the entire APOGEE data set of $\approx250,000$ spectra in ten minutes on a single, low-cost GPU. We release the stellar parameters and 18 individual-element abundances with associated uncertainty for the entire APOGEE DR14 dataset. Simultaneously, we release \texttt{astroNN}, a well-tested, open-source python package developed for this work, but that is also designed to be a general package for deep learning in astronomy. \texttt{astroNN} is available at \url{https://github.com/henrysky/astroNN} with extensive documentation at \url{http://astroNN.readthedocs.io}.

\end{abstract}





\section{Introduction}


With astronomy becoming increasingly characterized by large surveys and big data sets, machine-learning techniques have become staples of astronomical data analysis that are used for low-level data processing, classification, interpolation, pattern recognition, and parameter inference. Among modern machine-learning techniques, deep learning using artificial neural networks (ANNs) is getting increasing attention from astronomers, because of its great potential for data-driven astronomy and its recent successes in many fields such as computer vision, voice recognition, machine translation, etc. While ANNs have been around for decades, they have only become one of the dominant machine-learning techniques in the last few years. The reasons behind this recent progress are the combination of big data, cheap availability of fast computational hardware, advances in the methodology of ANNs, and the availability and accessibility of software platforms implementing this technology \citep{2017arXiv171205855S}. In astronomy, the big data era is now fully upon us: large photometric \citep[e.g., SDSS;][]{2011ApJS..193...29A}, spectroscopic \citep[e.g., APOGEE;][]{2017AJ....154...94M}, astrometric \citep[Gaia;][]{2016A&A...595A...1G}, and time-domain \citep[e.g., PTF;][]{2009PASP..121.1395L} data sets already exist and will grow exponentially in terms of quality and quantity with upcoming projects like the Large Synoptic Survey Telescope \citep[LSST;][]{2009arXiv0912.0201L}, Euclid \citep{2011arXiv1110.3193L}, and projects like the Maunakea Spectroscopic Explorer \citep[MSE;][]{2016arXiv160600043M}. ANNs are poised to play an outsized role in this data-rich future.

The hardware and software landscape for machine learning has vastly changed in the last decade. In particular, the availability of cheap graphics processing units (GPUs) is driven by the development and demand of high performance low cost personal gaming, but modern machine learning methods are ideally suited to be run on GPUs. For example, the code and analysis that we describe in this paper is entirely run on a personal desktop computer with a $\approx\$ 500$ consumer GPU accelerated by the NVIDIA CUDA Deep Neural Network library \citep{2014arXiv1410.0759C}. The accessibility of software technology for machine learning is supported by open source communities. For example, the open source python deep learning libraries used in this work---\texttt{Tensorflow} \citep{2016arXiv160304467A}, \texttt{keras} \citep{keras2015}, and the package developed by us and described in this paper \texttt{astroNN} \footnote{\url{https://github.com/henrysky/astroNN}} (see Appendix \ref{appendix:graph:astroNN}). The combination of these factors allows astronomers to easily exploit deep learning in astronomical big data analysis without the high cost of development in human resources, time, hardware, and software.

In this work we investigate the application of deep learning to the analysis of high-resolution spectroscopic data. Such data contain a wealth of information about the overall physical state of stars and about the abundances of different elements in their photospheres \citep{Gray05a}. This information is traditionally extracted using tools such as the curve of growth, equivalent widths, or forward modeling with synthetic spectra in what is often a laborious and tedious effort. Here we demonstrate that, as long as a small---thousands of stars---training set of data analyzed with more traditional means is available, ANNs can process high-resolution spectra faster and more reliably than other methods.

ANNs have been used for spectrosopic analysis before. Because of the paucity of high-resolution spectra until recently, early use of ANNs was mostly limited to training the network on libraries of synthetic spectra \citep[e.g.,][]{1997MNRAS.292..157B,2000A&A...357..197B,2015MNRAS.452..158Y}. The large spectroscopic databases provided by the SEGUE \citep{2009AJ....137.4377Y} and APOGEE surveys allowed ANNs to be trained directly on observed stellar spectra and map them onto stellar parameters (SEGUE: \citealt{2007A&A...467.1373R}, \citealt{2008AJ....136.2022L}; APOGEE: \citealt{2018MNRAS.475.2978F}). Uncertainty estimation has been explored using generative ANNs (GANs) for the \emph{Gaia} RVS data by \citet{2016A&A...594A..68D}. 

The availability of the large and rich APOGEE spectroscopic data set has spurred many applications of machine-learning methods to these data. Examples of these are the \texttt{Cannon 2} \citep{2016arXiv160303040C} method for data-driven abundance analysis, dimensionality reduction of the spectral space to determine the dimensionality of abundance space in the Milky Way \citep{2018MNRAS.475.1410P}, and machine-learned outlier detection and similarity directly using the spectra \citep{2018MNRAS.476.2117R}. 

The work most directly related to that described in this paper is the recent \texttt{StarNet} ANN, which is trained on spectroscopic data from APOGEE \citep{2018MNRAS.475.2978F}. \texttt{StarNet} uses a convolutional neural network to infer three labels $[\teff, \logg, \xh{Fe}]$ from high-resolution spectra and demonstrated that deep learning is an effective way both in terms of performance and of accuracy to do spectroscopic analysis when the number of training data is large.

In this work, we go beyond the \texttt{StarNet} method in various way: (a) we present a robust objective function for the neural network to learn from incomplete data while taking uncertainty in the training labels into account, (b) we use a Bayesian neural network with dropout variational inference with this objective function to estimate the uncertainties of labels determined by the neural network \citep{2015arXiv150602142G}, (c) we simultaneously infer 22 stellar and elemental abundance labels accurately and precisely for both high and low signal-to-noise ratio (SNR) spectra while constraining the model to reflect our physical understanding of stellar spectra, (d) we implement the method on a GPU using standard tools allowing for more than an order of magnitude speed-up and make these easily accessible (see Appendix \ref{subsec:fastMC}), and (e) we demonstrate that a large neural network can work well with a limited amount of training data (thousands of high SNR stellar spectra). We also present these 22 stellar parameters and elemental abundance predictions with uncertainties for the entire APOGEE DR14 data set.

The outline of this paper is as follows. Section \ref{sec:nn-general} describes the basics of ANNs, of dropout variational inference, and of our robust objective function. Section \ref{subsec:apogee-data} discusses the data selection and processing from APOGEE DR14 to construct training and test sets. Section \ref{sec:main_nn} describes the performance of our trained neural network on unseen individual and combined spectra in the test sets, on stars in open and globular clusters, and we present the results from a sensitivity analysis to understand the neural network. Section 5 describes variations in NN training such as: training on the full, uncensored spectrum, training with small data sets with only thousands of spectra, and training with a different continuum normalization process. Section 6 discusses the fast performance of the neural network and what types of future work this allows, and comparisons to other, similar spectroscopic analysis approaches. Section 7 gives our conclusions. Appendix A describes the \texttt{astroNN} python package developed for this work and gives instructions on how to perform variational inference on arbitrary APOGEE spectra.

Code to reproduce all of the plots in this paper as well as the \texttt{FITS}\footnote{\url{https://github.com/henrysky/astroNN_spectra_paper_figures/raw/master/astroNN_apogee_dr14_catalog.fits}} data file containing our neural network's predictions for 22 stellar parameters and abundances for the whole APOGEE DR14 is available at \url{https://github.com/henrysky/astroNN_spectra_paper_figures}.

\section{Bayesian Neural Networks with Drop-out}\label{sec:nn-general}

Deep learning refers to the usage of multi-layer (``deep'') ANNs to achieve both supervised and unsupervised machine learning. As opposed to task-specific algorithms, ANNs provide a general learning method that can be used on a variety of learning tasks. Bayesian neural network refers to the application of Bayesian inference to neural networks to obtain a posterior distribution function (PDF) on the weights that characterize the ANN given some input data. This PDF can then be used to propagate training uncertainty into predictions made with the ANN using new input data. Here, we use dropout variational inference as an approximation to Bayesian neural networks. Dropout variational inference is a new method proposed by \citet{2015arXiv150602142G} that can be applied to a wide variety of neural network architectures, is easy to implement, and is computationally cheap. This technique is previously used in \citet{2017ApJ...850L...7P} for strong gravitational lensing parameters estimation with uncertainty

In this section, we give a brief introduction to ANNs, describe the idea of dropout variational inference, and then discuss the loss function that we use to train our ANN using incomplete and noisy training data.

\subsection{Artificial Neural Networks}

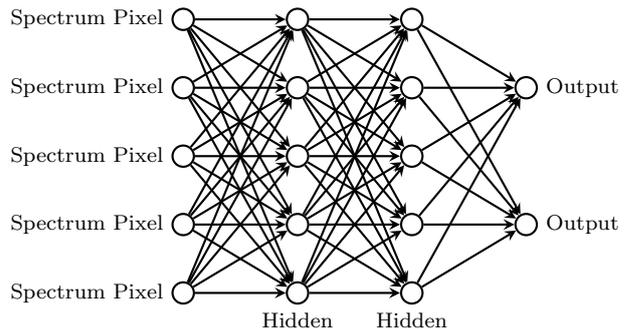
\begin{figure}
\centering
\begin{tikzpicture}

	\node[circle, draw, thick, label=left:{Spectrum Pixel}] (i1) {};
	\node[circle, draw, thick, above=2em of i1, label=left:{Spectrum Pixel}] (i2) {};
	\node[circle, draw, thick, above=2em of i2, label=left:{Spectrum Pixel}] (i3) {};
	\node[circle, draw, thick, below=2em of i1, label=left:{Spectrum Pixel}] (i4) {};
	\node[circle, draw, thick, below=2em of i4, label=left:{Spectrum Pixel}] (i5) {};
	
	\node[circle, draw, thick, right=4em of i1] (h1) {};
	\node[circle, draw, thick, right=4em of i2] (h2) {};
	\node[circle, draw, thick, right=4em of i3] (h3) {};
	\node[circle, draw, thick, right=4em of i4] (h4) {};
	\node[circle, draw, thick, right=4em of i5, label=below:{Hidden}] (h5) {};
	
	\node[circle, draw, thick, right=4em of h1] (hh1) {};
	\node[circle, draw, thick, right=4em of h2] (hh2) {};
	\node[circle, draw, thick, right=4em of h3] (hh3) {};
	\node[circle, draw, thick, right=4em of h4] (hh4) {};
	\node[circle, draw, thick, right=4em of h5, label=below:{Hidden}] (hh5) {};
	
	\node[circle, draw, thick, right=4em of hh2, label=right:{Output}] (o1) {};
	\node[circle, draw, thick, right=4em of hh4, label=right:{Output}] (o2) {};
	
	\draw[-stealth, thick] (i1) -- (h1);
	\draw[-stealth, thick] (i1) -- (h2);
	\draw[-stealth, thick] (i1) -- (h3);
	\draw[-stealth, thick] (i1) -- (h4);
	\draw[-stealth, thick] (i1) -- (h5);
	\draw[-stealth, thick] (i2) -- (h1);
	\draw[-stealth, thick] (i2) -- (h2);
	\draw[-stealth, thick] (i2) -- (h3);
	\draw[-stealth, thick] (i2) -- (h4);
	\draw[-stealth, thick] (i2) -- (h5);
	\draw[-stealth, thick] (i3) -- (h1);
	\draw[-stealth, thick] (i3) -- (h2);
	\draw[-stealth, thick] (i3) -- (h3);
	\draw[-stealth, thick] (i3) -- (h4);
	\draw[-stealth, thick] (i3) -- (h5);
	\draw[-stealth, thick] (i4) -- (h1);
	\draw[-stealth, thick] (i4) -- (h2);
	\draw[-stealth, thick] (i4) -- (h3);
	\draw[-stealth, thick] (i4) -- (h4);
	\draw[-stealth, thick] (i4) -- (h5);
	\draw[-stealth, thick] (i5) -- (h1);
	\draw[-stealth, thick] (i5) -- (h2);
	\draw[-stealth, thick] (i5) -- (h3);
	\draw[-stealth, thick] (i5) -- (h4);
	\draw[-stealth, thick] (i5) -- (h5);
	
	\draw[-stealth, thick] (h1) -- (hh1);
	\draw[-stealth, thick] (h1) -- (hh2);
	\draw[-stealth, thick] (h1) -- (hh3);
	\draw[-stealth, thick] (h1) -- (hh4);
	\draw[-stealth, thick] (h1) -- (hh5);
	\draw[-stealth, thick] (h2) -- (hh1);
	\draw[-stealth, thick] (h2) -- (hh2);
	\draw[-stealth, thick] (h2) -- (hh3);
	\draw[-stealth, thick] (h2) -- (hh4);
	\draw[-stealth, thick] (h2) -- (hh5);
	\draw[-stealth, thick] (h3) -- (hh1);
	\draw[-stealth, thick] (h3) -- (hh2);
	\draw[-stealth, thick] (h3) -- (hh3);
	\draw[-stealth, thick] (h3) -- (hh4);
	\draw[-stealth, thick] (h3) -- (hh5);
	\draw[-stealth, thick] (h4) -- (hh1);
	\draw[-stealth, thick] (h4) -- (hh2);
	\draw[-stealth, thick] (h4) -- (hh3);
	\draw[-stealth, thick] (h4) -- (hh4);
	\draw[-stealth, thick] (h4) -- (hh5);
	\draw[-stealth, thick] (h5) -- (hh1);
	\draw[-stealth, thick] (h5) -- (hh2);
	\draw[-stealth, thick] (h5) -- (hh3);
	\draw[-stealth, thick] (h5) -- (hh4);
	\draw[-stealth, thick] (h5) -- (hh5);

	\draw[-stealth, thick] (hh1) -- (o1);
	\draw[-stealth, thick] (hh1) -- (o2);
	\draw[-stealth, thick] (hh2) -- (o1);
	\draw[-stealth, thick] (hh2) -- (o2);
	\draw[-stealth, thick] (hh3) -- (o1);
	\draw[-stealth, thick] (hh3) -- (o2);
	\draw[-stealth, thick] (hh4) -- (o1);
	\draw[-stealth, thick] (hh4) -- (o2);
	\draw[-stealth, thick] (hh5) -- (o1);
	\draw[-stealth, thick] (hh5) -- (o2);
\end{tikzpicture}
\caption{A simple multilayer neural network.}
\label{figure:ann}
\centering
\end{figure}

ANNs were originally inspired by biological systems such as human brains, which consist of numerous neurons interconnected by synapses. The information or stimuli from the external world travel in the form of electrical impulses called action potentials. An ANN mimics this configuration and dynamics by representing a general learning task as a set of layers consisting of neurons that communicate through connections (the ``synapses''). The ``strength'' of each connection is given by a simple linear functional form $y = w\,x+b$, which is transformed by each neuron using a non-linear function. The final decision or output of the ANN depends on the input and on the strength of the connections (the weights $\vec{W}$). There is no need to pre-program any knowledge in an ANN, i.e., neural networks consist of random weights at the initial training stage. In order to achieve learning, error signals representing the agreement between the true output and the ANN output for (a subset of) the training set is back-propagated through the neural network and the connection strength of the synapses is adjusted to obtain better agreement between truth and prediction. Figure \ref{figure:ann} shows an example of a typical ANN.

Mathematically, an ANN is a real-valued, smooth function approximation to a general function. Consider data $\{\vec{x}, \vec{y}\}$ and a neural network $f$ parameterized by a set of parameters $\vec{W}$ that takes input $\vec{x}$ and maps it to $\hat{\vec{y}}=f^{\vec{W}}(\vec{x})$. Each neuron $i$ in an ANN takes an input vector $\vec{v}$ (either the actual input $\vec{x}$ or the output of a previous layer) and maps it to an output number $o$ as $o = \vec{w}_i \,\vec{v}+b_i$; this output number is then optionally transformed using a non-linear function before going to the next layer or the output. The parameters $\vec{W}$ in our notation represent the total set of $\{\vec{w}_i,b_i\}$ of all the neurons. 

The quality of the ANN is represented by an objective function $J(\vec{y}, \hat{\vec{y}})$ that we want to minimize. At each step in the training process, a new set of parameters $\vec{W}_{\mathrm{new}}$ in ANN optimization can be obtained by descending along the gradient computed using back-propagation \citep{1986Natur.323..533R}
\begin{equation} \label{eq:grad_descent}
\vec{W}_\mathrm{new} = \vec{W} - \eta\frac{\partial{J}}{\partial{\vec{W}}}\,,
\end{equation}
where $\eta$ is the learning rate. For small $\eta$, this update step should lead to $J(\vec{y}, f^{\vec{W}_{\mathrm{new}}}(\vec{x}))$ that is smaller than $J(\vec{y}, f^\vec{W}(\vec{x}))$. To deal with large data sets, the gradients in each step are typically computed using only a small, random subset of the training data set that is different in each update step; this corresponds to a \emph{stochastic gradient descent} algorithm. In our work, we use a more sophisticated version of this type of optimizer, the ADAM optimizer \citep{2014arXiv1412.6980K}.

``Convolutional neural networks'' (CNNs) refers to the method of learning a set of convolution filters as part of the learning process. These convolution filters act as feature extractors, by convolving the input data (or the input data to a given layer in the ANN) with the filter. Useful features are extracted after these convolutional layers and, because they usually are not densely packed in the input data, we can apply a technique called max-pooling to reduce the size of our neural network, thus prevent overfitting. Max-pooling of size $n$ takes the maximum of $n$ pixels in non-overlapping subregions of incoming data, thus reducing the size of the input. Another advantage of max-pooling for spectroscopic analysis is that it induces a degree of translational invariance, making the analysis independent of small errors in the radial velocity correction.

\subsection{Dropout Variational Inference}

Variational inference is a general Bayesian inference method where the PDF is obtained not by sampling---as is the case when one uses Markov Chain Monte Carlo (MCMC) methods---but rather by fitting an approximation to the PDF using an objective function obtained by variational calculus. This method of Bayesian inference has the advantage that it can be applied to problems with large numbers of parameters, because optimization is in general faster and easier than sampling. This advantage comes at the expense of accuracy: the obtained PDF is only an approximation to the true PDF.

A Bayesian neural network with dropout variational inference works by approximating the true PDF for the weights ---which can number in the millions in our application below ---as a product of Bernoulli distributions. It can then be shown that a neural network trained with dropout applied to every layer except the last one and which has a Gaussian prior on the weights is an approximation to the full Bayesian neural network \citep{2015arXiv150602142G}. The Gaussian prior is achieved by imposing L2 regularization, parameterized by a regularization constant $\lambda$, on the loss function as
\begin{equation} \label{eq:l2_reg}
J_\mathrm{regularized}(\vec{y}, \hat{\vec{y}}) = J(\vec{y}, \hat{\vec{y}}) + \lambda\vec{W}^2\,.
\end{equation}
The hyper-parameter $\lambda$ is determined using the validation set: We set $\lambda$ to the value that optimizes the neural network precision on a validation set. L2 regularization is equivalent to a Gaussian prior in the Bayesian interpretation.

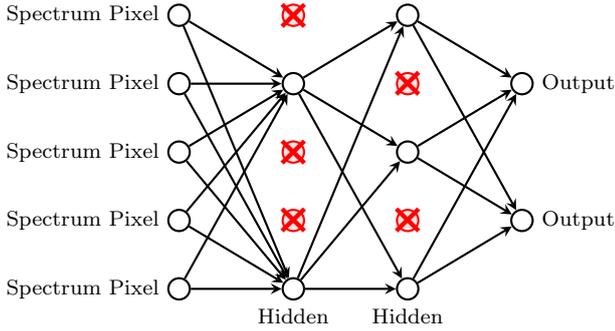
\begin{figure}
\centering
\begin{tikzpicture}

	\node[circle, draw, thick, label=left:{Spectrum Pixel}] (i1) {};
	\node[circle, draw, thick, above=2em of i1, label=left:{Spectrum Pixel}] (i2) {};
	\node[circle, draw, thick, above=2em of i2, label=left:{Spectrum Pixel}] (i3) {};
	\node[circle, draw, thick, below=2em of i1, label=left:{Spectrum Pixel}] (i4) {};
	\node[circle, draw, thick, below=2em of i4, label=left:{Spectrum Pixel}] (i5) {};
	
	\node[circle, draw, thick, red, fill=red!10, right=4em of i1] (h1) {};
	\node[circle, draw, thick, right=4em of i2] (h2) {};
	\node[circle, draw, thick, red, fill=red!10, right=4em of i3] (h3) {};
	\node[circle, draw, thick, red, fill=red!10, right=4em of i4] (h4) {};
	\node[circle, draw, thick, right=4em of i5, label=below:{Hidden}] (h5) {};
	
	\node[red] (icr) at (h1) {$\mathlarger{\mathlarger{\mathlarger{\mathlarger{\mathlarger{\bm{\times}}}}}}$};
	\node[red] (icr) at (h3) {$\mathlarger{\mathlarger{\mathlarger{\mathlarger{\mathlarger{\bm{\times}}}}}}$};
	\node[red] (icr) at (h4) {$\mathlarger{\mathlarger{\mathlarger{\mathlarger{\mathlarger{\bm{\times}}}}}}$};
	
	\node[circle, draw, thick, right=4em of h1] (hh1) {};
	\node[circle, draw, thick, red, fill=red!10, right=4em of h2] (hh2) {};
	\node[circle, draw, thick, right=4em of h3] (hh3) {};
	\node[circle, draw, thick, red, fill=red!10, right=4em of h4] (hh4) {};
	\node[circle, draw, thick, right=4em of h5, label=below:{Hidden}] (hh5) {};
	
	\node[red] (icr) at (hh2) {$\mathlarger{\mathlarger{\mathlarger{\mathlarger{\mathlarger{\bm{\times}}}}}}$};
	\node[red] (icr) at (hh4) {$\mathlarger{\mathlarger{\mathlarger{\mathlarger{\mathlarger{\bm{\times}}}}}}$};
	
	\node[circle, draw, thick, right=4em of hh2, label=right:{Output}] (o1) {};
	\node[circle, draw, thick, right=4em of hh4, label=right:{Output}] (o2) {};

    \draw[-stealth, thick] (i1) -- (h2);
    \draw[-stealth, thick] (i1) -- (h5);
    \draw[-stealth, thick] (i2) -- (h2);
    \draw[-stealth, thick] (i2) -- (h5);
	\draw[-stealth, thick] (i3) -- (h2);
	\draw[-stealth, thick] (i3) -- (h5);
	\draw[-stealth, thick] (i4) -- (h2);
	\draw[-stealth, thick] (i4) -- (h5);
	\draw[-stealth, thick] (i5) -- (h2);
	\draw[-stealth, thick] (i5) -- (h5);
	
	\draw[-stealth, thick] (h2) -- (hh1);
	\draw[-stealth, thick] (h2) -- (hh3);
	\draw[-stealth, thick] (h2) -- (hh5);
	\draw[-stealth, thick] (h5) -- (hh1);
	\draw[-stealth, thick] (h5) -- (hh3);
	\draw[-stealth, thick] (h5) -- (hh5);
	
	\draw[-stealth, thick] (hh1) -- (o1);
	\draw[-stealth, thick] (hh1) -- (o2);
	\draw[-stealth, thick] (hh3) -- (o1);
	\draw[-stealth, thick] (hh3) -- (o2);
	\draw[-stealth, thick] (hh5) -- (o1);
	\draw[-stealth, thick] (hh5) -- (o2);

\end{tikzpicture}
\caption{An example of dropout: a certain fraction of neurons are randomly dropped when evaluating the ANN. This is primarily used to prevent overfitting, but can also provide uncertainty estimation when it is used as an approximation to a Bayesian neural network.}
\label{figure:ann_dropout}
\centering
\end{figure}

Dropout \citep{2012arXiv1207.0580H} is a technique that is primarily used to avoid over-fitting in neural networks, because deep neural networks usually have more parameters than data points. Dropout multiplies the hidden neurons---that is, those not in the input or output layer---by a Bernoulli distributed random variable that take the value 1 with a certain probability and 0 otherwise. When neurons are multiplied by zero they are effectively dropped; doing this during training prevents neurons from co-adapting to the training data, which would otherwise lead to overfitting. An example of a given instance of dropout for the example ANN in Figure \ref{figure:ann} is shown in Figure \ref{figure:ann_dropout}.

To use dropout for uncertainty estimation, we run $N$ times Monte Carlo dropout in forward passes through the network; in other words, we keep dropout turned on to make predictions using the ANN. Since dropout drops weights randomly, the neural network is probabilistic and has different predictions in every forward pass through the network. The mean value of predictions will be the final prediction and the standard deviation of predictions will be the model uncertainty. In addition to this  ``model uncertainty'', the neural network that we use also gives a ``predictive uncertainty'' (see below for how we obtain the predictive uncertainty). The total uncertainty is the sum of model and predictive uncertainty in quadrature \citep{2017arXiv170304977K}.

In more mathematical terms, the Bayesian neural network predicts $\{\hat{\vec{y}}, \hat{\vec{\sigma}}^2\}=f^{\widehat{\vec{W}}}(x)$ where $\hat{\vec{y}}$ is the prediction of the labels, $\hat{\vec{\sigma}}^2$ is the predictive variance, $f^{\widehat{\vec{W}}}$ is the neural network with randomly masked weights due to dropout and $\vec{x}$ is the input data. We run the forward pass in the neural network   $N$ times and obtain a set of $\{\hat{\vec{y}}_i, \hat{\vec{\sigma}}^2_i\}^N_{i=1}$. The final prediction $\hat{\vec{y}}$ and uncertainty intervals $\hat{\vec{\sigma}}$ is
\begin{equation} \label{eq:1sigma}
\hat{\vec{y}}\pm\hat{\vec{\sigma}} = \frac{1}{N}\sum^{N}_{i=1}\hat{\vec{y}}_i\pm\sqrt{\frac{1}{N}\sum^{N}_{i=1}\hat{\vec{y}}^2_i - \left(\frac{1}{N}\sum^{N}_{i=1}\hat{\vec{y}}_i \right)^2 + \frac{1}{N}\sum^{N}_{i=1}\hat{\vec{\sigma}}^2_i}\,.
\end{equation}

\subsection{Objective function for incomplete and noisy training data} \label{subsec:objective}

The objective function $J(\vec{y}, \hat{\vec{y}})$ is the function that the neural network aims to minimize to train the network. Generally, neural networks for regression use the Mean Squared Error (MSE), defined as
\begin{equation} \label{eq:mse}
\text{Mean Squared Error} = J_{MSE}(\vec{y}, \hat{\vec{y}}) = \frac{1}{N} \sum^N_{i=1}(\hat{\vec{y}_i}-\vec{y}_i)^2\,.
\end{equation}
The MSE is prone to overfitting to outliers due to the squared term and it does not take uncertainty in the training labels into account.

However, astronomical data and observations are often incomplete and noisy. For example, in the case of spectroscopic data, the abundance of an element $\xh{X_1}$ may have only a single absorption line in the wavelength range by the detector. But due to reasons such as cosmic rays or if the line's Earth-frame wavelength happens to overlap with a strong sky emission line, the only absorption line $\xh{X_1}$ may not be measurable. But  other abundances $\xh{X_2}$ can still be accurately measured. Previous data-driven approaches on spectroscopic data like  the \texttt{Cannon 2} \citep{2016arXiv160303040C} or \texttt{StarNet} \citep{2018MNRAS.475.2978F} needed to filter such spectra from the training set because of the data incompleteness.

For the Bayesian neural network used in this work, we employ the following robust objective to get the loss for label $i$ and assume that unavailable data are labeled using \texttt{MAGIC NUM}
\begin{equation} \label{eq:mmse}
   J(y_i, \hat{y}_i) =
   \begin{cases}
            \frac{1}{2} (\hat{y_i}-y_i)^2 e^{-s_i} + \frac{1}{2}(s_i) & \text{ for } y_i \neq \texttt{MAGIC NUM}\\
            0 & \text{ for } y_i = \texttt{MAGIC NUM}
   \end{cases}
\end{equation}

In this expression, $s_i = \ln \left[\sigma^2_{\mathrm{known}, i} + \sigma^2_{\mathrm{predictive}, i}\right]$, which corresponds to the natural logarithm of the sum of the known uncertainty variance in the labels and an additional predictive variance. This predictive variance is also learned by the neural network and forms another output from the neural network for each input. The known uncertainty variance is that returned by the reduction pipeline that produces the labels for the training subset. In general, the neural network can be trained to give the predictive variance without any known variance in the labels, which is a form of unsupervised training. The predictive variance from the loss function learned by the neural network represents any variance in the training set that cannot be explained by the known variance. This predictive variance contributes to the error budget for predictions on new data, see Equation \eqref{eq:1sigma}.

The final loss for the stochastic gradient descent is calculated from a mini-batch partition of the data consisting of $N$ data point and $D$ labels
\begin{equation}  \label{eq:fcorrect}
J(\vec{y}, \hat{\vec{y}}) = \frac{1}{N} \sum_{i=1}^{N} \left( \frac{1}{D} \sum_{i=1}^{D} J(y_i, \hat{y}_i)  \right)\mathcal{F}_{\mathrm{correction},i}\,,
\end{equation}
where $\mathcal{F}_{\mathrm{correction},i}$ is a correction term to correct for the fact that in Equation \eqref{eq:mmse} we effectively assume that the neural network made no error for missing data. If $D$ is the overall number of labels and $D_i$ is the number of labels not equal to \texttt{MAGIC NUM} for data point $i$, then
\begin{equation}  \label{eq:fcorrect_itself}
\mathcal{F}_{\mathrm{correction},i} = \frac{D}{D_i}\,.
\end{equation}

The objective function in Equation \eqref{eq:mmse} acts as a robust version of the conventional MSE objective in Equation \eqref{eq:mse}, allowing the neural network to take the effect of uncertain and missing labels into account. This makes the model more robust because high uncertainty in a training label will have a smaller effect on the loss, preventing the neural network to learn from such labels. Equation \eqref{eq:mmse} also assumes that, without any other information, the prediction from the neural network is accurate and thus back-propagates zero loss for incomplete labels. The correction term $\mathcal{F}_{\mathrm{correction},i}$ in  Equation \eqref{eq:fcorrect} is factored into the final loss in order to prevent the effective learning rate from decreasing due to the presence of missing labels. $\mathcal{F}_{\mathrm{correction},i}$ equals one in which there are no missing labels, in which case it resembles a conventional loss function. 

\section{High-resolution spectroscopic data from APOGEE}\label{subsec:apogee-data}

\begin{figure}
\centering
\includegraphics[width=0.5\textwidth]{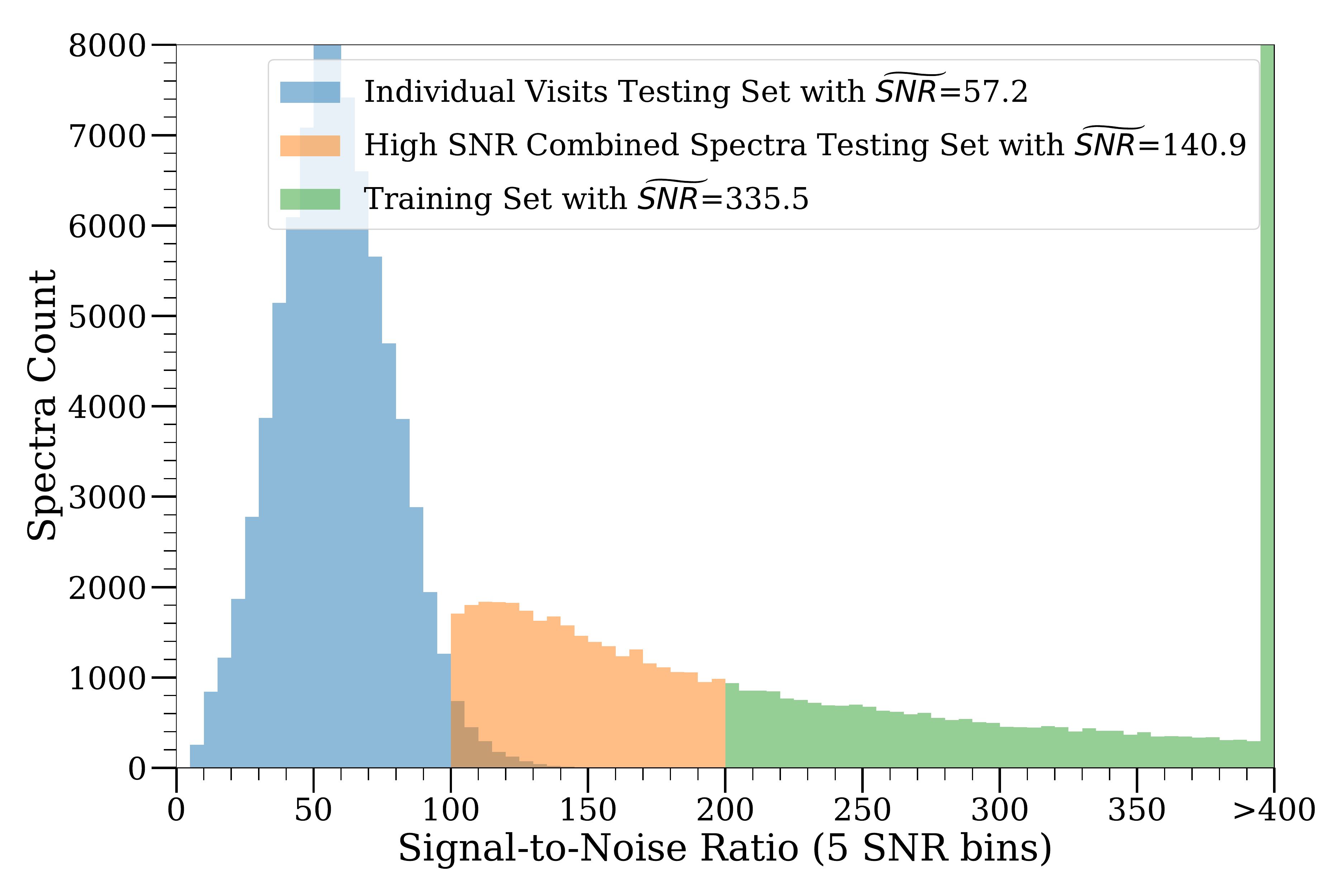}
\caption{Signal-to-noise (SNR) ratio distribution in the training set and in the two test sets used in this work. $\widetilde{SNR}$ represents median SNR. All spectra with ASPCAP reported SNR$>400$ are set to SNR$=400$, leading to the peak at high SNR. High SNR combined spectra test set refers to combined spectra with $100<\mathrm{SNR}<200$. Individual visits test set refers to the set of individual visits of stars in the high SNR combined-spectra test set.}
\label{figure:snr_traintest}
\centering
\end{figure}

The main source of data that we use are spectra and derived labels from the APO Galactic Evolution Experiment (APOGEE) Data Release 14 \citep{2018arXiv180709773H,2018arXiv180709784J,2018ApJS..235...42A}. APOGEE spectra are obtained with a 300-fibre spectrograph \citep{Wilson10a} attached to the Sloan Foundation 2.5m telescope at Apache Point Observatory \citep{2006AJ....131.2332G}. APOGEE is an infrared ($1.5 \mu m - 1.7 \mu m$), high resolution ($R\sim 22,500$), high signal-to-noise ratio (typical $\mathrm{SNR}>100$) spectroscopic survey. The APOGEE data set contains stellar parameter and chemical abundances obtained using the APOGEE Stellar Parameter and Chemical Abundances Pipeline (ASPCAP; \citealt{2016AJ....151..144G}). ASPCAP is an automated pipeline for determining the stellar labels from observed spectra by comparing observed spectra to a precomputed library of theoretical spectra using $\chi^2$ minimization. We describe how we select data from the overall APOGEE DR14 catalog and how we define our training and tests sets in Section \ref{subsec:selection}. In Section \ref{subsec:reduction} we discuss the method to process the data in the  training and test sets.

\subsection{Training, test, and validation data selection from APOGEE DR14} \label{subsec:selection}

\begin{figure}
\centering
\includegraphics[width=0.5\textwidth]{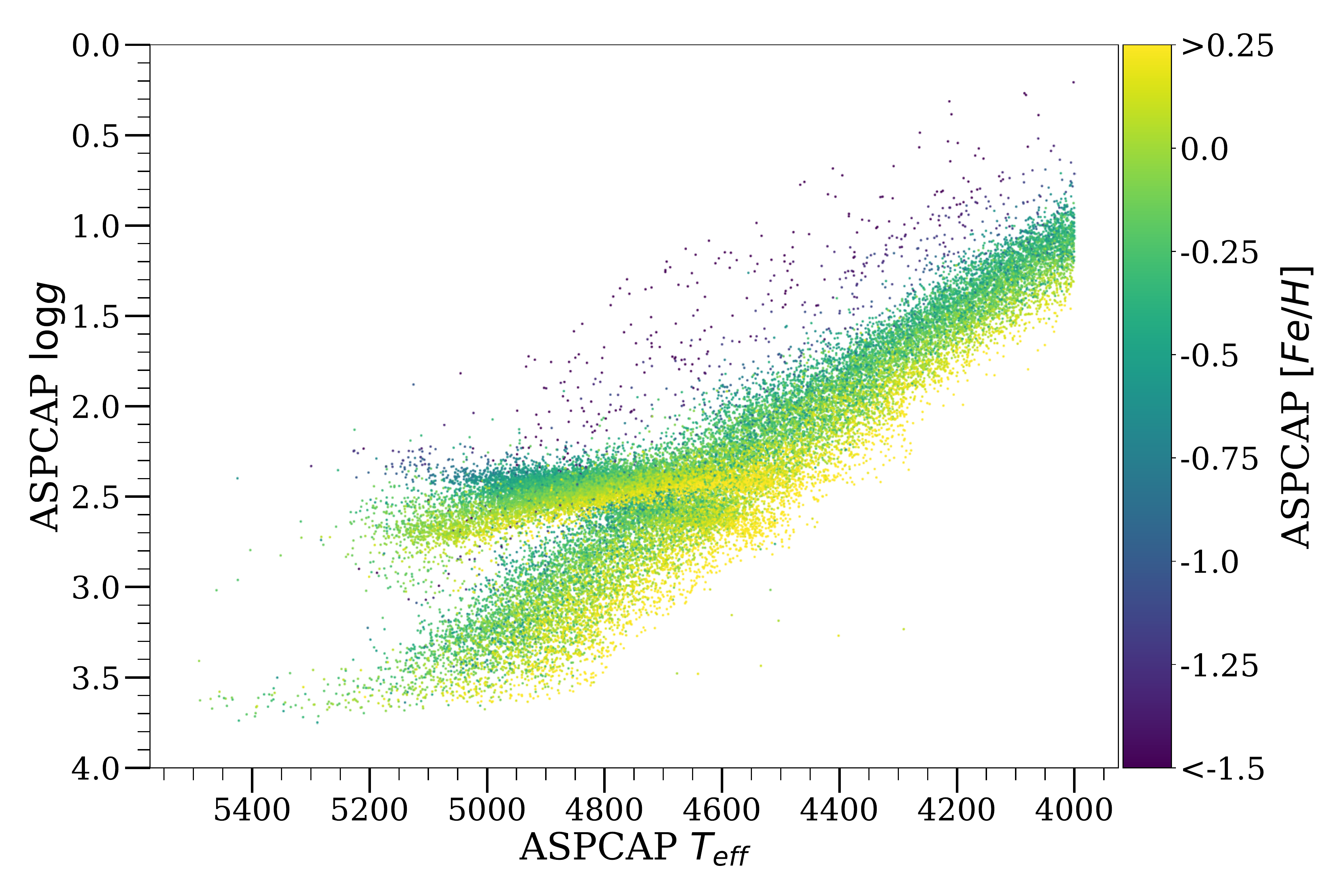}
\caption{$\logg$ vs $\teff$ colored by $\xh{Fe}$ abundances in the training set; all the labels in the training set are determined by ASPCAP. All main-sequence stars in APOGEE DR14 have $\logg$ set to \texttt{MAGIC NUM} $= -9999$ and are not displayed in the plot.}
\label{figure:train_tefflogg}
\centering
\end{figure}

We have created one training and two test sets from the set of APOGEE DR14 spectra. Each data set consists of continuum normalized spectra, 22 ASPCAP labels (\teff, \logg, \xh{C}, \xh{CI}, \xh{N}, \xh{O}, \xh{Na}, \xh{Mg}, \xh{Al}, \xh{Si}, \xh{P}, \xh{S}, \xh{K}, \xh{Ca}, \xh{Ti}, \xh{TiII}, \xh{V}, \xh{Cr}, \xh{Mn}, \xh{Fe}, \xh{Co}, \xh{Ni})\footnote{\xh{CI} and \xh{TiII} are measurements of the carbon and titanium abundance using spectral regions that only have neutral atomic carbon and singly-ionized atomic titanium features, respectively. The overall \xh{C} and \xh{Ti} are mostly determined by molecular (for carbon) and neutral atomic (for titanium) features. Because ASPCAP returns these measurements separately, we include them also as part of our label set. The masking windows used can be accessed with the python function described in \url{https://astronn.readthedocs.io/en/v1.0.0/tools_apogee.html\#retrieve-aspcap-elements-window-mask}.} and their associated ASPCAP uncertainty. ASPCAP determines the abundances of all of these elements using synthetic spectra computed using the line list from \citet{2015ApJS..221...24S}. The SNR distributions of these subsets are shown in Figure \ref{figure:snr_traintest}. In Figure \ref{figure:train_tefflogg} we display the training sample in the space of \teff\ and \logg\ colored by \xh{Fe}. The training set consists of 33,407  spectra that have $\mathrm{SNR}>200$. At the start of training, $90\%$ of the training set is randomly selected to train the neural network---that is, used to compute the gradients of the objective function in the training steps---and the remaining $10\%$ constitute a separated validation set used to validate the performance of the neural network during the training process. $4.6\%$ of the combination of all training ASPCAP labels are $-9999$, the value used by ASPCAP to represent highly uncertain or unavailable labels; as discussed in Section \ref{subsec:objective}, these spectra are still used in our robust objective function (thus, \texttt{MAGIC NUM} $= -9999$ for APOGEE in Equation [\ref{eq:mmse}]). Two test sets are used, one consists of spectra with SNR between 100 to 200, which are called the high-SNR test set. These spectra are picked from the set of ``combined'' APOGEE spectra, which are the combinations of the individual exposures and it is these combinations that are used by ASPCAP for their analysis (most APOGEE spectra are obtained as a set of at least three individual hour-long exposures to obtain $\mathrm{SNR}>100$; \citealt{2015AJ....150..148H}). This set of spectra is entirely separate from the training set. The other test set consists of 81,483 spectra picked from the set of individual visit spectra that go into the combined spectra in the high-SNR test set. These spectra in the individual visit test set have much lower SNR than the spectra in the set of combined training spectra (which all have $\mathrm{SNR}>200$), see Figure \ref{figure:snr_traintest}. The advantage of this test set is that we are interested in the performance of our neural network for spectra with low SNR, but ASPCAP does not provide labels for individual-visit spectra with low SNR. However, we do have neural network or ASPCAP labels for the high-SNR combinations of the individual visit spectra, which we can use to test our method. Because the low SNR test set consists of the same \emph{stars} as the high SNR test set, the labels in the high SNR test set are representative of those in the low SNR test set, even though the \emph{noise} in them is not. Thus, the low SNR test set provides a stringent test of how this method performs at low SNR.

On top of the SNR cut, we perform cuts on the values of the stellar parameters. This is necessary, because all of the knowledge learned by the neural network is solely driven by the training data. Therefore, we need to make sure that the training labels are as accurate as possible, because any systematic inaccuracy such as bias at lower SNR will be captured by the neural network and propagated to new test data. For this reason, we exclude spectra with surface temperature $\teff$ smaller than $4000\,\mathrm{K}$ or higher than $5500\,\mathrm{K}$, because at these temperatures ASPCAP may not be accurate \citep{2016AJ....151..144G}. Additionally, we remove spectra flagged with the \texttt{APOGEE\_ASPCAPFLAG} or \texttt{APOGEE\_STARFLAG} flags and spectra with a radial velocity scatter larger than $1\,\mathrm{km\,s}^{-1}$, because these represent potential issues with specific labels, issues with spectra, and potential binary stars, respectively. These cuts ensure the quality of the training and test set.

ASPCAP determines abundances by performing a $\chi^2$ fit on an interpolated, large grid of synthetic spectra computed for the APOGEE wavelength region \citep{2016AJ....151..144G}. After performing the fit, ASPCAP made several calibrations of the stellar parameters and abundances based on the consistency of abundances within open and globular clusters and by comparing to external data. The synthetic spectra are computed under various simplifying assumptions and using line lists that are not 100\,\% complete and this limits the quality of the synthetic grid used by ASPCAP and thus ultimately limits the Neural Network accuracy. Any systematic bias will propagated from ASPCAP to Neural Network during training. This is a disadvantage of the specific supervised machine-learning method that we are using here (we discuss advantages and disadvantages of our method compared to other methods in more detail in Sections \ref{subsec:assumption} and \ref{subsec:comparesynth}).

The resulting training set contains both main-sequence dwarfs and red giants. However, ASPCAP does not report calibrated values of \logg\ for main-sequence stars and these are all set to \texttt{MAGIC NUM} $=-9999$. These \logg\ labels are therefore ignored in our training procedure, although other labels such as \teff\ for main-sequence are used. This means that we cannot hope to determine good \logg\ for main-sequence stars with our neural network. We discuss this further below.

\subsection{Data reduction for training} \label{subsec:reduction}

\begin{figure}
\centering
\includegraphics[width=0.5\textwidth]{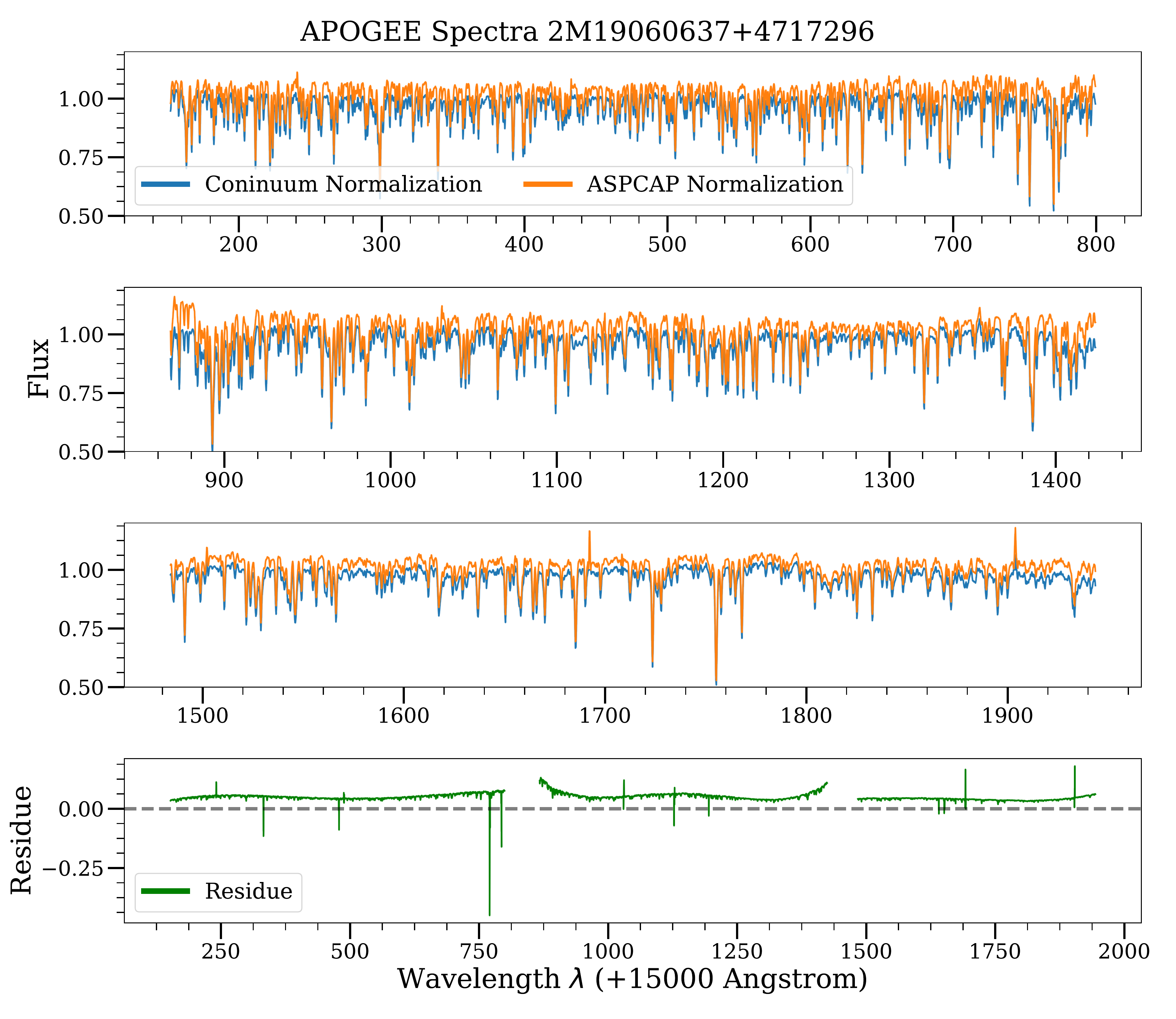}
\caption{Example of continuum normalization: this Figure shows the continuum-normalized combined APOGEE spectrum of 2M19060637+4717296, showing the difference between the continuum-normalization performed by ASPCAP spectra and and that used in our method. The top three panels display the spectrum in three parts (corresponding to the three detectors used in the APOGEE instrument), while the bottom panels shows the difference between the two. The difference between the two normalization methods is a relatively-smooth function of wavelength and there is an overall offset due to the fact that ASPCAP does not attempt to trace the actual continuum, while our method does.}
\label{figure:norm_aspcap}
\centering
\end{figure}

All methods for spectroscopic analysis work with continuum-normalized spectra, because the information about stellar labels is mainly contained in narrow spectral features and the overall continuum is typically not well calibrated. The method used for continuum normalization is important, because we require a method that is invariant with respect to SNR due to the fact that we train the neural network on high SNR spectra only but also test it on low SNR spectra. To accomplish this, we use the same method as employed by the \texttt{Cannon 2} method \citep{2016arXiv160303040C}, where continuum normalization is performed by using a set of ``continuum pixels'' ---identified as pixels that depend little on the stellar labels using their data-driven model for APOGEE spectra---and fitting a continuum to the flux in only these pixels. The spectrum in each APOGEE detector is normalized separately in this method, with a 2-degree polynomial which has been demonstrated to be effective by \citet{2016arXiv160303040C}. While we use DR14 spectra, we use a set of continuum pixels obtained using DR12 spectra that is included in the \texttt{apogee} software package \citep{2016ApJ...817...49B}. Spectra in DR14 extend over a slightly wider wavelength range for each detector than those in DR12, which means that our continuum normalization does not perfectly capture the behavior of the continuum at the edges of the detector. But as we show below, the exact method of continuum normalization does not have a large effect on the performance of our method. 

Figure \ref{figure:norm_aspcap} shows the difference between the continuum normalization used by ASPCAP and that described above for the combined spectrum of 2M19060637+4717296 as an example. An ideal continuum normalization would place continuum pixels to have a flux of 1. It is clear that the ASPACP normalization fails to do so (note that this is by design: the DR14 ASPCAP analysis explicitly does not attempt to do this; see \citealt{2018arXiv180709773H}). Aside from an overall offset, the continuum-normalized spectra are similar for both methods.

After continuum normalization, we check the APOGEE pixel-level mask bits in the \texttt{APOGEE\_PIXMASK} bitmask and set the flux value of pixels that contain the following bits to $1$ (the expected continuum): \textbf{0}: bad pixel, \textbf{1}: cosmic ray, \textbf{2}: saturated, \textbf{3}: unfixable, \textbf{4}: bad from dark, \textbf{5}: bad from flat, \textbf{6}: high error, \textbf{7}: no sky info, \textbf{12}: overlaps a significant sky line.

Besides the continuum normalization of the spectra, we need to standardize the labels and continuum normalized spectra to be able to easily use them with standard neural-network methods. For the labels, we subtract the mean and divide by the standard deviation such that the training labels all approximately have a mean of 0 and a standard deviation of 1. In our case we have 22 labels, and therefore we calculate 22 means and 22 standard deviations to standardize the labels. Labels which are \texttt{MAGIC NUM} $=-9999$ (that is, missing) are not involved in the normalization process, i.e. $-9999$ values remain constant such that objective function in Equation \eqref{eq:mmse} can recognize these missing labels during optimization. For continuum-normalized spectra, we calculate the mean of the flux pixel-by-pixel for the spectra in training set, and subtract the means from all the spectra in the training set. In other words ideally an average spectrum should be a flat straight line at $\mathrm{flux}=0$ after this final normalization step. The reason behind this normalization is that this maps the average spectrum to the average label for a neural network with all weights set to zero and we expect that the average spectrum should have stellar labels close to the average of those in the training set. Thus, it will be easier and faster for the neural network to converge to a minimum.

While performing variational inference on test sets, it is important to use the same normalization procedure and the same set of means and standard deviations as used for the training set to normalize and denormalize all the data. This is also the reason why we do not scale the training spectra to have standard deviation of $1$, because the training and testing spectra have completely different SNR. In other words, training set spectra have high SNR, thus low overall standard deviation and test set spectra have low SNR, thus higher overall standard deviation. Using the parameters to standardize training spectra will not standardize testing spectra so we chose not to scale any spectra.

\section{Performance on APOGEE data} \label{sec:main_nn}

\begin{figure}
\centering
\includegraphics[width=0.5\textwidth, clip]{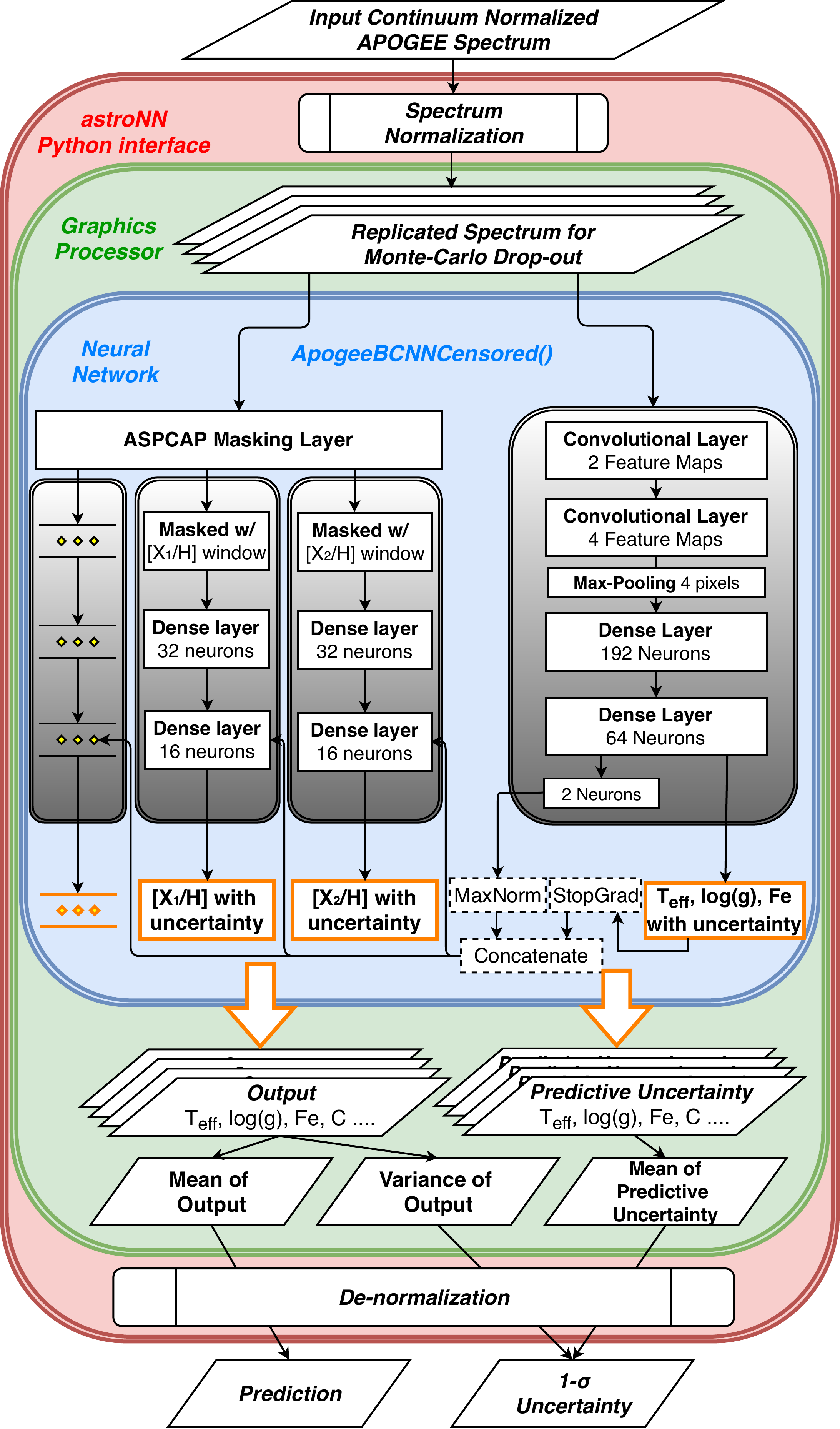}
\caption{The neural-network architecture mainly used in this work, defined as \texttt{ApogeeBCNNCensored()} in \texttt{astroNN}.}
\label{figure:nn_flow}
\centering
\end{figure}

\begin{figure*}
\centering
\includegraphics[width=\textwidth]{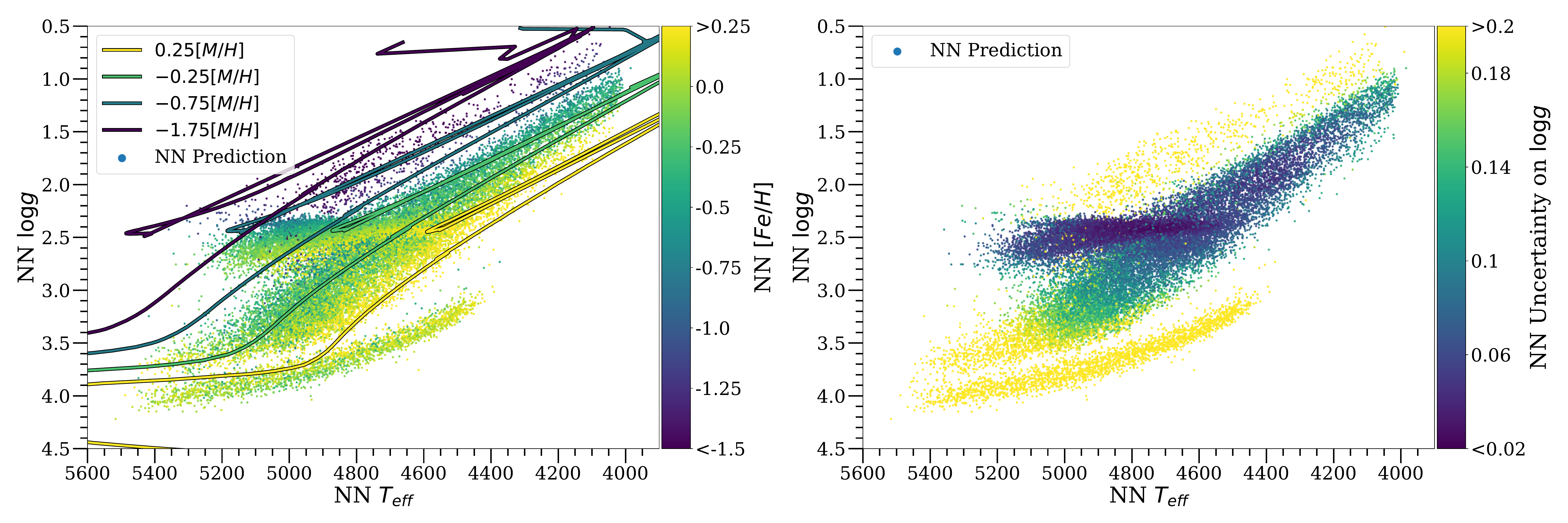}
\caption{Neural Network $\teff$ and $\logg$ prediction color-coded by $\xh{Fe}$ (left panel) and $\logg$ uncertainties (right panel). The group at high $\logg$ is due to 2,611 spectra in the high-SNR test set that are dwarfs with \texttt{MAGIC NUM} $=-9999$ ASPCAP value. \logg\ values for these stars are absent in the training data,  because ASPCAP in DR14 does not provide the $\logg$ for dwarfs. The neural network clearly predicts the wrong values for \logg\, but this is also reflected in the large uncertainties for these stars in the right panel. For test set objects along the giant branch, the neural network returns reasonable parameters (compare to Figure~\ref{figure:train_tefflogg}). Similarly, the uncertainty in \logg\ is high for low-metallicity, low \logg\ giants, for which training data are sparse.}
\label{figure:logg_teff_isochrones}
\centering
\end{figure*}

The main neural network that we train and test with APOGEE spectra in this work is the network \verb|ApogeeBCNNCensored()| in \texttt{astroNN} (see Appendix~\ref{appendix:graph:astroNN}). The neural network architecture is shown in Figure \ref{figure:nn_flow}. Users only have to provide a continuum-normalized spectrum and are returned a prediction and associated uncertainty. Rather than using a single, simple neural network to predict stellar parameters and elemental abundances from an input spectrum, we use a combination of (a) a large neural network trained on the full spectral wavelength range to predict [$\teff$, $\logg$, $\xh{Fe}$] (the big gray-colored network on the right side in Figure \ref{figure:nn_flow}) and (b) 19 mini neural networks used to predict the 19 \xh{X} abundances based on fixed regions of the spectrum that contain known spectral features for each element and the overall [$\teff$, $\logg$, $\xh{Fe}$]. This architecture mimics that of traditional spectroscopic analysis and of ASPCAP, where the overall stellar parameters ($\teff, \logg, \xh{Fe}$, etc.) are determined first and individual abundances are determined afterwards from specific spectral features. However, our method differs from this in a crucial aspect: the network trained to predict [$\teff$, $\logg$, $\xh{Fe}$] from the full spectrum also has a two-neuron connection characterized by two latent variables to the 19 mini-networks used to predict the individual abundances (in addition to feeding [$\teff$, $\logg$, $\xh{Fe}$] to the mini-networks as well). We choose to use two neurons as the connection to mimic ASPCAP, in which \xfe{C} and $\xfe{\alpha}$ are fitted to the full spectrum, because these elements strongly affect the stellar photosphere and thus the formation of all spectral lines. By using two neurons we can learn a latent space similar to these two elements, but we do not require the latent variables to exactly correspond to \xfe{C} and $\xfe{\alpha}$ to give the network the opportunity to learn a better low-dimensional set of latent variables. This allows the mini-networks to use a limited amount of information from the full spectrum that is not captured by [$\teff$, $\logg$, $\xh{Fe}$] in making their predictions. To produce the 19 masked spectra for the 19 mini-networks (one mask per element) we use the windows employed by ASPCAP DR14 to determine individual abundances \citep{2016AJ....151..144G}. These windows were derived by the ASPCAP team using synthetic spectral syntheses to isolate regions of the spectra most sensitive to individual elements. Unlike ASPCAP, we only use the windows as a binary mask---pixels are either in or out---we do not use the weights assigned to pixels within the windows. 

The entire combination of the large ANN to predict [$\teff$, $\logg$, $\xh{Fe}$] and the 19 mini-networks, including their two-neuron connection, is trained simultaneously. To achieve this, two unconventional layers are included in the network, \texttt{StopGrad} and \texttt{MaxNorm}. \texttt{StopGrad} is an identity transformation layer with the property that its gradient is always set to 0 during training, but otherwise is simply 1 (for example, when computing the sensitivity of the network output to input; see Section \ref{subsec:jac}). This layer prevents the error from an individual abundance $\xh{X}$ to be back-propagated during training to the network predicting [$\teff$, $\logg$, $\xh{Fe}$]. That is, we do not allow prediction errors in $\xh{X}$ to affect the training of the network that predicts the stellar parameters.  \texttt{MaxNorm} is a weight-constraint layer that requires $\sqrt{\sum{w^2}} \leq \delta$, where $w$ are the weights and $\delta$ is a constraint constant. $\delta$ is determined using the validation data set, such that $\delta$ optimizes the performance of neural network on the validation set. This layer prevents the mini neural networks that predict $\xh{X}$ from paying too much attention to the full spectrum. The reason why we construct this rather complex network rather than the more straightforward single network that predicts all the labels from the full spectrum is discussed in detail in Section \ref{subsec:full_spec}. Briefly, the reason is that when allowing the full spectrum to inform individual abundances, the ANN predictions in regions of label-space with few training data become highly correlated due to correlations in the training data. 

We discuss the performance of the \verb|ApogeeBCNNCensored()| network in detail in the following subsections. We will see that we find that the neural network trained on high SNR training data and tested on high SNR testing data displays a fairly high bias, which may result from errors in ASPCAP, but a small amount of scattering. When testing on low SNR individual-visit spectra, the network shows almost no bias and small amounts of scattering (but larger than that at higher SNR). The results are considerably better than any previous work on applying machine-learning techniques to high-resolution spectral analysis. The uncertainty estimation from dropout variational inference that we find makes sense, because it correlates with $1 / \sqrt{SNR}$ and is similar to the scatter in the residuals between our method and ASPCAP. For both test sets, we find that the NN performs the best for elements that ASPCAP reports as being their most accurate elements (for example \xh{Mg} and \xh{Ni} in an independent validation of ASPCAP DR13/14 by \citet{2018arXiv180709784J}). A sensitivity analysis of how the ANN outputs depend on the input spectra shows that the model depends in a reasonable manner on wavelength. All our results can be reproduced using our online code (see Appendix~\ref{appendix:graph:astroNN}), but due to the stochastic nature of dropout and the neural-network training process, it is impossible to reproduce the exact same results. However, statistically, the results should be very close to those described in this work.

In the following, we define the residual as 
\begin{equation}  \label{eq:residual}
\mathrm{Residual} = \textit{NN } \mathrm{Prediction}-\mathrm{ASPCAP}\,,
\end{equation}
where $NN$ refers to the neural network. As a robust measurement of the scatter, we use a measure based on the Median Absolute Deviation (MAD): $\madstd = 1.4826\,\mathrm{MAD}$, where the factor is such that for a Gaussian distribution $\madstd$ equals the Gaussian standard deviation. Thus, for a set of residual $R:[R_1, R_2,...,R_n]$, \madstd\ is
\begin{equation}  \label{eq:scattering}
\madstd = 1.4826\,\mathrm{median}\left(|R_i - \mathrm{median}(R)|\right)\,.
\end{equation}
In all of these calculations, ASPCAP labels which are equal to $-9999$ (or, more generally, labels equal to \texttt{MAGIC NUM}) are excluded from the calculation.

\subsection{Comparison to ASPCAP at high signal-to-noise ratio}\label{subsec:highsnr}

\begin{table}
  \centering
  \caption{Neural-network prediction result on the high SNR test set from comparing NN predictions to ASPCAP}
  \label{table:highsnr_result}
	\begin{tabular}{lrr} 
		\hline
        Label & Median of residual & \madstd\ of residual\\
		\hline
        $\teff$ & $-20 \text{ K}$  & $30 \text{ K}$ \\
        $\logg$ & $0.012 \text{ dex}$  & $0.051 \text{ dex}$ \\
        $\xh{C}$ & $0.003 \text{ dex}$ & $0.040 \text{ dex}$ \\
        $\xh{CI}$ & $0.013 \text{ dex}$  & $0.058 \text{ dex}$ \\
        $\xh{N}$ & $-0.004 \text{ dex}$  & $0.041 \text{ dex}$ \\
        $\xh{O}$ & $-0.021 \text{ dex}$  & $0.046 \text{ dex}$ \\
        $\xh{Na}$ & $-0.01 \text{ dex}$  & $0.16 \text{ dex}$ \\
        $\xh{Mg}$ & $0.000 \text{ dex}$  & $0.027 \text{ dex}$ \\
        $\xh{Al}$ & $-0.038 \text{ dex}$ & $0.071 \text{ dex}$ \\
        $\xh{Si}$ & $0.000 \text{ dex}$  & $0.029 \text{ dex}$ \\
        $\xh{P}$ & $-0.02 \text{ dex}$  & $0.10 \text{ dex}$ \\
        $\xh{S}$ & $0.006 \text{ dex}$  & $0.051 \text{ dex}$ \\
        $\xh{K}$ & $-0.013 \text{ dex}$  & $0.049 \text{ dex}$ \\
        $\xh{Ca}$ & $-0.015 \text{ dex}$  & $0.033 \text{ dex}$ \\
        $\xh{Ti}$ & $-0.029 \text{ dex}$  & $0.052 \text{ dex}$ \\
        $\xh{TiII}$ & $0.06 \text{ dex}$  & $0.17 \text{ dex}$ \\
        $\xh{V}$ & $-0.009 \text{ dex}$  & $0.097 \text{ dex}$ \\
        $\xh{Cr}$ & $-0.002 \text{ dex}$  & $0.048 \text{ dex}$ \\
        $\xh{Mn}$ & $-0.018 \text{ dex}$  & $0.038 \text{ dex}$ \\
        $\xh{Fe}$ & $-0.004 \text{ dex}$  & $0.020 \text{ dex}$ \\
        $\xh{Co}$ & $-0.02 \text{ dex}$  & $0.14 \text{ dex}$ \\
        $\xh{Ni}$ & $0.003 \text{ dex}$  & $0.029 \text{ dex}$ \\
		\hline
	\end{tabular}
\end{table}

\begin{figure*}
\centering
\includegraphics[width=\textwidth]{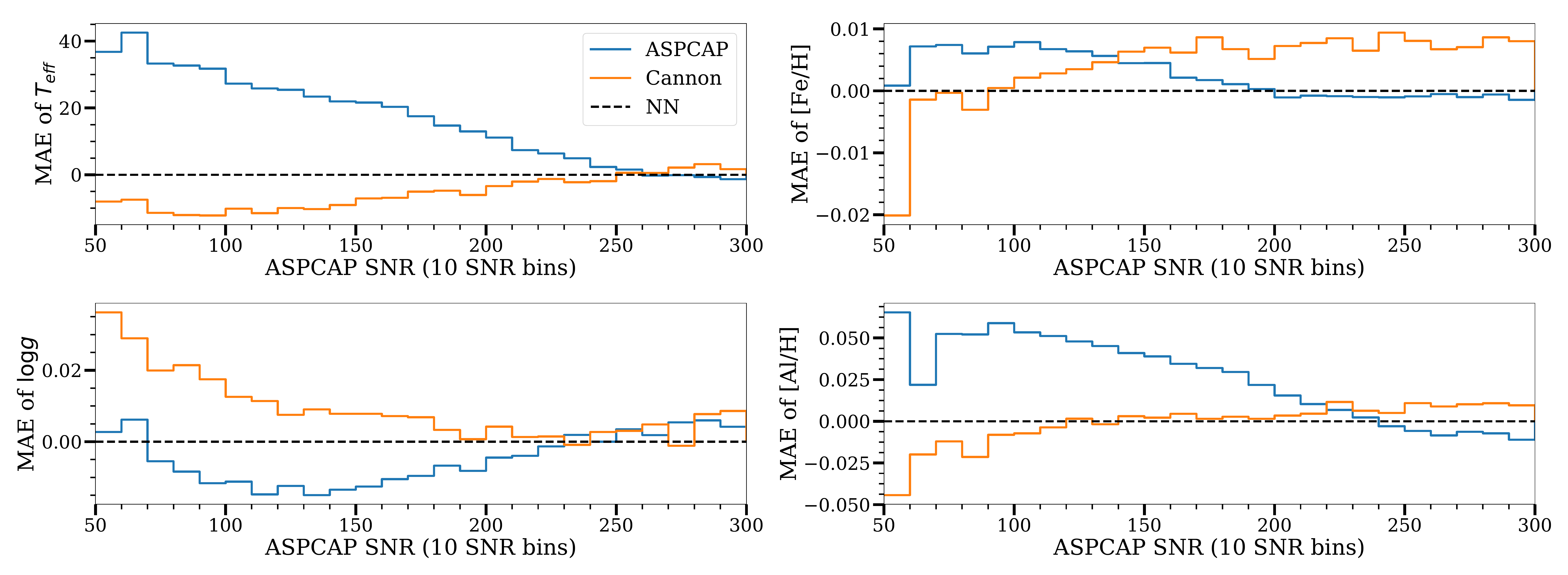}
\caption{Median offset between our NN method, ASPCAP, and the \texttt{Cannon 2}. Each panel shows the Median Absolute Error (MAE) of ASPCAP or the \texttt{Cannon 2} assuming that the NN is the ground truth in bins of SNR for the combined APOGEE spectra. The panels show the MAE of \teff, \logg, and \xh{Fe}, as well as \xh{Al} as a representative element. The MAE bias between the NN and ASPCAP has a strong SNR dependence for \teff\ and somewhat less strong for \logg. This trend is much smaller between the NN and the \texttt{Cannon 2}, especially for \xh{Al}. Typically, the bias between ASPCAP and the \texttt{Cannon 2} (the difference between the blue and orange curves) is higher than that between either of them and the NN. The strong trend with SNR when comparing the NN to the \texttt{Cannon 2} or to ASPCAP may be due to a SNR dependent bias in ASPCAP.}
\label{figure:lowSNR_issue}
\centering
\end{figure*} 

\begin{figure*}
\centering
\includegraphics[width=\textwidth]{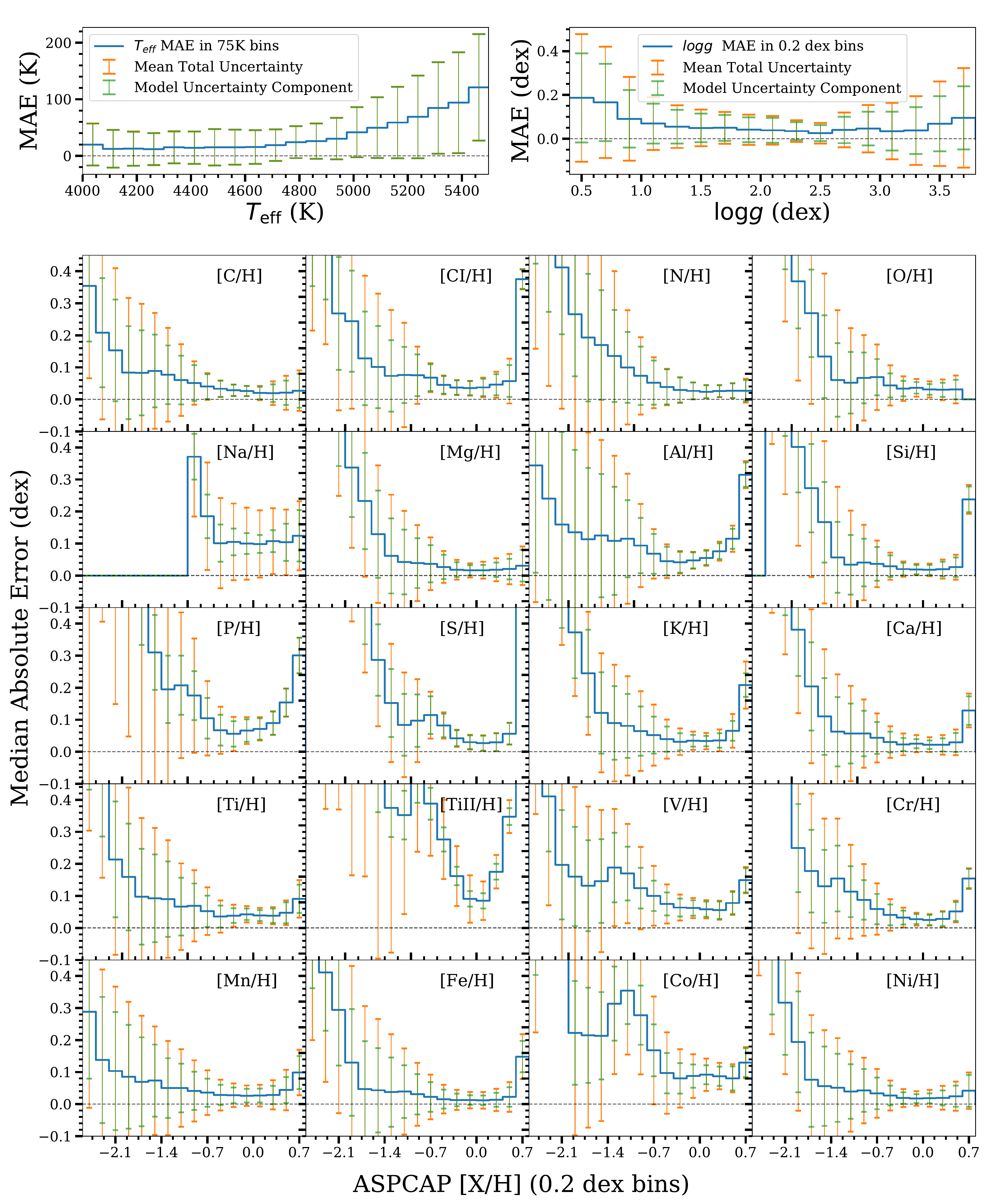}
\caption{Comparison between neural-network predictions for $\teff$, $\logg$, and $\xh{X}$ and those from ASPCAP at high SNR (SNR between 100 and 200). The blue curve shows the MAE between the NN and ASPCAP in bins of the ASPCAP label, while the green and orange error bars give the NN model and total uncertainty. Overall the MAE is small and the uncertainties are similar to the MAE, but there are bigger residuals at low \xh{X}, because the training set contains few low-metallicity stars and spectral features are weaker for such stars, leading to worse performance of the NN.}
\label{figure:all_you_can_residue_highSNR}
\centering
\end{figure*}

\begin{figure*}
\centering
\includegraphics[width=\textwidth]{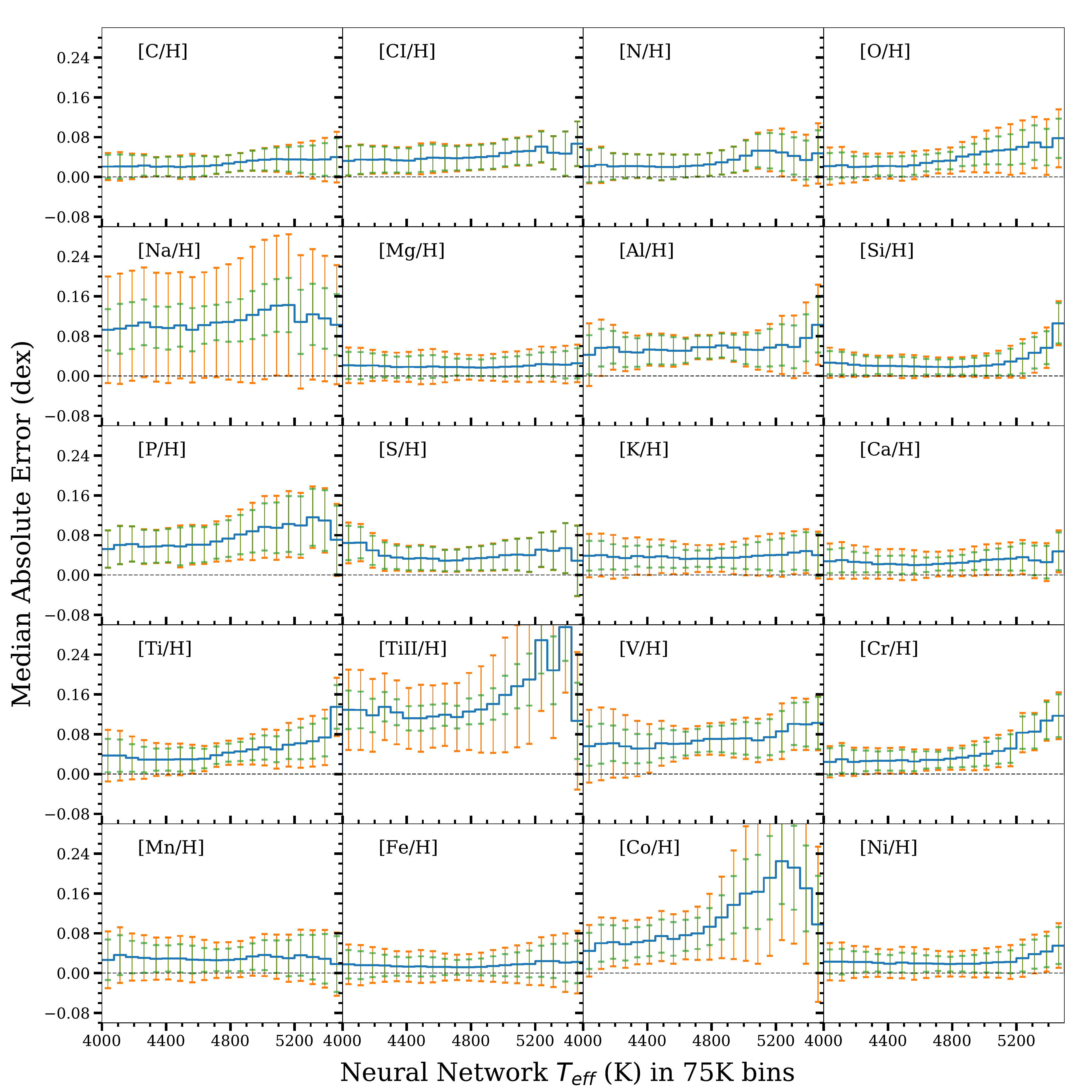}
\caption{Comparison between neural-network predictions for $\xh{X}$ and the ASPCAP labels for the high SNR test set as a function of \teff.  Curves and errorbars are as in Figure \ref{figure:all_you_can_residue_highSNR}. Spectral features are weaker at higher \teff\ leading to an increase in the typical residuals that is matched by an increase in the NN uncertainty.}
\label{figure:delta_xh_teff_highSNR}
\centering
\end{figure*}

We first test the performance of the neural network for high SNR spectra, which we define here as spectra with SNR between $100$ to $200$ (the standard cuts described in Section \ref{subsec:selection} apply as well). In this section, we evaluate the performance of the NN by comparing it to the predictions from ASPCAP for the same stars. Thus, any comparison is affected by biases in the ASPCAP results themselves and we will see that there are reasons to believe that ASPCAP suffers from SNR-dependent biases, even at these high SNRs. These biases inflate the size of the residuals between the neural-network predictions and ASPCAP. We also do not account for the random uncertainties in the ASPCAP predictions, which also contribute to the size of the residuals. In this sense, the biases and errors derived from comparing to ASPCAP from this section are an upper limit on the size of the biases and errors of the neural network's predictions. In the next section, we will compare the neural-network's prediction against itself and find much smaller residuals. 

A summary of the results is given in Table \ref{table:highsnr_result}. Figure \ref{figure:logg_teff_isochrones} displays the predicted $\logg$ versus $\teff$. In this figure, the left panel is color-coded by $\xh{Fe}$ and we overlay PARSEC isochrones \citep{2012MNRAS.427..127B} at 4 different metallicities to indicate the expected location of stars in this space. It is clear that the predicted parameters for stars at different metallicities conform well to these expectations. That the predicted \logg\ is highly precise for giants is clear from the narrowness of the red clump. The right panel of Figure \ref{figure:logg_teff_isochrones} shows the same predicted $\logg$ versus $\teff$, but now color-coded by the neural-network uncertainty in \logg. The neural network is highly confident in its \logg\ prediction along the well-populated parts of the giant branch. However, the uncertainty in \logg\ is large for the group of stars at high \logg. This is reasonable, because these are main-sequence dwarfs for which we have no \logg\ training data (see discussion in Section \ref{subsec:selection} above). The predicted \logg\ are clearly wrong, but this is reflected in the high uncertainties for these stars. Similarly, the uncertainty in \logg\ is high for low-metallicity, low-\logg\ giants, for which training data are sparse (see Figure~\ref{figure:train_tefflogg}).

\begin{figure*}
\centering
\includegraphics[width=\textwidth]{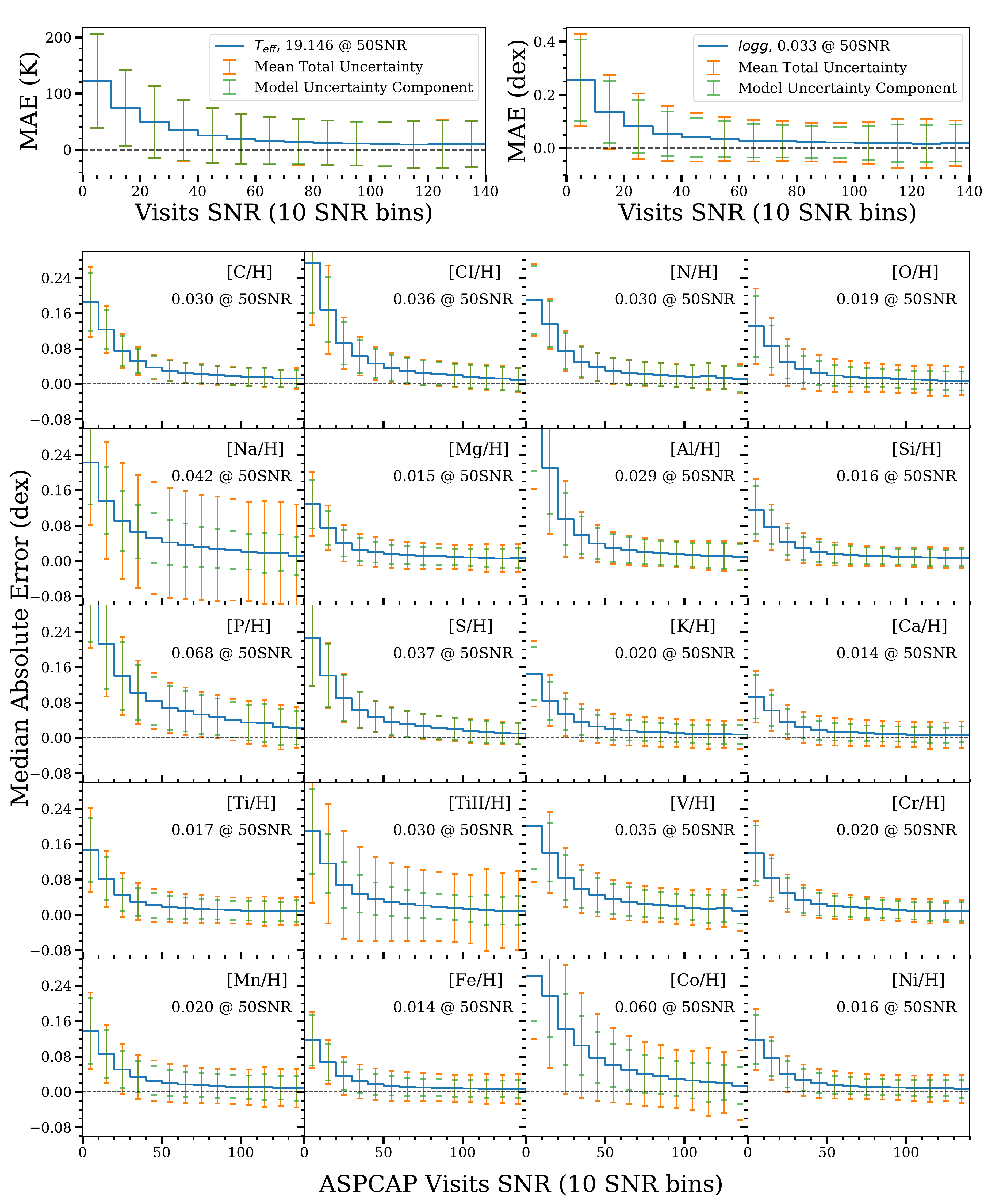}
\caption{Comparison between neural-network predictions for $\teff$, $\logg$, and $\xh{X}$ from low-SNR, individual-exposure spectra and those from the same neural network applied to the combined spectra in high-SNR test set. Curves and errorbars are as in Figure \ref{figure:all_you_can_residue_highSNR}. The number beside or below the parameter/abundance label in each plot represents the MAE at SNR$=50$. When compared to its own predictions at high SNR, the performance of the neural network at low SNR is excellent, with residuals of size $0.01 \text{ to } 0.02\,\mathrm{dex}$ at SNR$\gtrsim100$ and residuals that are only slightly bigger even at SNR$\approx50$.}
\label{figure:snr_acc}
\centering
\end{figure*}

The results in  Table \ref{table:highsnr_result} show that the neural network prediction displays a relatively high bias in all labels when compared to ASPCAP, especially in $\teff$. It is possible that the majority of this bias comes from the ASPCAP prediction on lower SNR spectra instead of the neural network. The neural network's predictions are consistent across a wide range of SNR, while the ASPCAP $\teff$ is probably biased at SNR$<200$. Figure \ref{figure:lowSNR_issue} shows the Median Absolute Error (MAE) of ASPCAP$-$NN and \texttt{Cannon}$-$NN between SNR 50 to 300 for the entire APOGEE DR14 catalog with good \texttt{APOGEE\_ASPCAPFLAG} and \texttt{APOGEE\_STARFLAG} flags, velocity scatter smaller than $1\,\mathrm{km\,s}^{-1}$, and ASPCAP $4000< \teff <5500$. Most labels show strong SNR-dependent trends that are above the empirical precision found for the neural network (see discussion below and Figure \ref{figure:snr_acc}). The bias with respect to ASPCAP is generally larger than that compared to the \texttt{Cannon 2}. These SNR dependent trends demonstrate that both ASPCAP and the \texttt{Cannon 2} are probably biased to a larger degree and at higher SNR than previously thought.

Figure \ref{figure:all_you_can_residue_highSNR} shows more detailed results on the performance of the neural network. In this figure, the blue line represents the median absolute error between the neural network and ASPCAP in bins of ASPCAP labels, the orange error bars represent the median total NN uncertainty in these bins, and the green line gives the contributions of the NN model uncertainty to the total uncertainty. The difference between the orange and green error bar therefore gives a sense of the contribution of the predictive uncertainty (the uncertainties are added in quadrature, so the predictive uncertainty is not simply the difference between the orange and green uncertainties). Both the accuracy and the precision are good for $\xh{X}\gtrsim-0.7$, especially around solar metallicity where there is the most training data. The neural network has lower accuracy at low $\xh{X}$ mainly due to two reasons: first, there are not much training data at low metallicity because $\xh{Fe}\approx -0.7$ is the lower bound of typical abundances in the Galactic disk and second, spectral features at low metallicity are less defined and therefore less informative about the abundances. The \logg\ prediction is accurate and precise in the well-populated lower and mid regions of the giant branch, but becomes more uncertain for very luminous, low-\logg\ giants. By and large, the uncertainties returned by the NN are similar to the typical MAE difference between the NN and ASPCAP. For some individual elements, the uncertainties are smaller than the typical MAE difference at low metallicity, but it is always the case that these uncertainties are quite large ($\gtrsim 0.25\,\mathrm{dex}$), thus signaling that the NN prediction is noisy.

Figure \ref{figure:delta_xh_teff_highSNR} similarly shows how the accuracy and uncertainties in the individual abundances depend on the surface temperature. Spectral features in stars with higher surface temperatures are weaker, so we expect the accuracy to decrease and the uncertainties to increase. Figure \ref{figure:delta_xh_teff_highSNR} demonstrates that this is indeed the case, although in general the trend with temperature is quite weak.

\subsection{Results at low signal-to-noise ratio} \label{subsec:lowsnr}

\begin{table}
  \centering
  \caption{Neural-network prediction result on the individual visits of low SNR test set spectra from comparing results on the combined spectra of the same test set}
    \label{table:indi_result}
	\begin{tabular}{lrr} 
		\hline
        Label & Median of residual & \madstd\ of residual\\
		\hline
        $\teff$ & $-3 \text{ K}$  & $29 \text{ K}$ \\
        $\logg$ & $0.000 \text{ dex}$  & $0.047 \text{ dex}$ \\
        $\xh{C}$ & $-0.004 \text{ dex}$ & $0.045 \text{ dex}$ \\
        $\xh{CI}$ & $-0.003 \text{ dex}$  & $0.054 \text{ dex}$ \\
        $\xh{N}$ & $-0.002 \text{ dex}$  & $0.046 \text{ dex}$ \\
        $\xh{O}$ & $0.000 \text{ dex}$  & $0.029 \text{ dex}$ \\
        $\xh{Na}$ & $-0.013 \text{ dex}$  & $0.060 \text{ dex}$ \\
        $\xh{Mg}$ & $-0.001 \text{ dex}$  & $0.023 \text{ dex}$ \\
        $\xh{Al}$ & $-0.005 \text{ dex}$ & $0.045 \text{ dex}$ \\
        $\xh{Si}$ & $-0.001 \text{ dex}$  & $0.024 \text{ dex}$ \\
        $\xh{P}$ & $0.00 \text{ dex}$  & $0.10 \text{ dex}$ \\
        $\xh{S}$ & $0.002 \text{ dex}$  & $0.054 \text{ dex}$ \\
        $\xh{K}$ & $-0.003 \text{ dex}$  & $0.030 \text{ dex}$ \\
        $\xh{Ca}$ & $-0.002 \text{ dex}$  & $0.021 \text{ dex}$ \\
        $\xh{Ti}$ & $-0.003 \text{ dex}$  & $0.026 \text{ dex}$ \\
        $\xh{TiII}$ & $-0.001 \text{ dex}$  & $0.043 \text{ dex}$ \\
        $\xh{V}$ & $-0.003 \text{ dex}$  & $0.052 \text{ dex}$ \\
        $\xh{Cr}$ & $-0.006 \text{ dex}$  & $0.029 \text{ dex}$ \\
        $\xh{Mn}$ & $-0.005 \text{ dex}$  & $0.030 \text{ dex}$ \\
        $\xh{Fe}$ & $-0.004 \text{ dex}$  & $0.020 \text{ dex}$ \\
        $\xh{Co}$ & $-0.010 \text{ dex}$  & $0.086 \text{ dex}$ \\
        $\xh{Ni}$ & $-0.005 \text{ dex}$  & $0.024 \text{ dex}$ \\
		\hline
	\end{tabular}
\end{table}

To test the neural network at lower SNR, we make use of the set of individual-exposures for the stars in the high-SNR test set, as explained in detail in Section \ref{subsec:selection}. For the combined spectra in the high-SNR test set, we have results from the neural network and we can therefore test the performance of the neural network on the low SNR individual exposures by comparing the predictions from the low SNR spectra to the results from the NN on the high SNR combined spectra.

The results from comparing the NN predictions on low SNR spectra to the NN measurements from their counterpart combined spectra are shown in Table \ref{table:indi_result}. The first thing to note is that the bias (median of the residual) with respect to the NN is much smaller than it was in the high SNR comparison in Table \ref{table:highsnr_result}. The main reason for this is that ASPCAP parameters and abundances are accurate at SNR$>200$ and the predictions from the neural network using the individual exposures are essentially the same parameters as the prediction from their high SNR counterpart. This demonstrates that the neural network's predictions are robust at low SNR.

\begin{figure}
\centering
\includegraphics[width=0.5\textwidth]{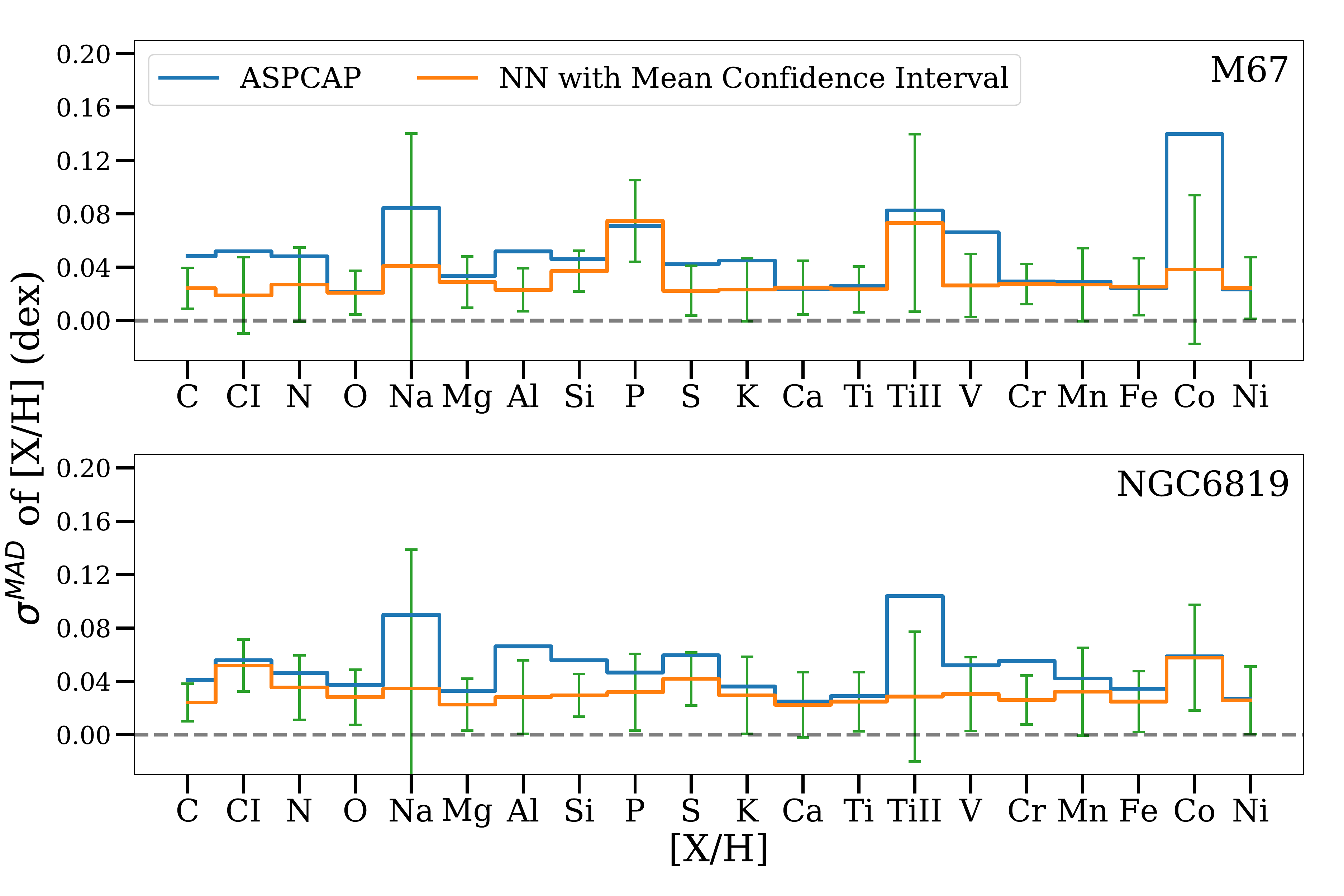}
\caption{Spread in the predicted NN and ASPCAP labels for stars in the open clusters M67 (14 stars) and NGC6819 (20 stars). The spread in the NN predictions is small for all elements and well matched by the NN uncertainties (green errorbars). The NN has a smaller spread in each element than ASPCAP. The overall level of chemical homogeneity is $0.030\pm0.029\,\mathrm{dex}$.}
\label{figure:open_clusters}
\centering
\end{figure}

\begin{figure}
\centering
\includegraphics[width=0.5\textwidth]{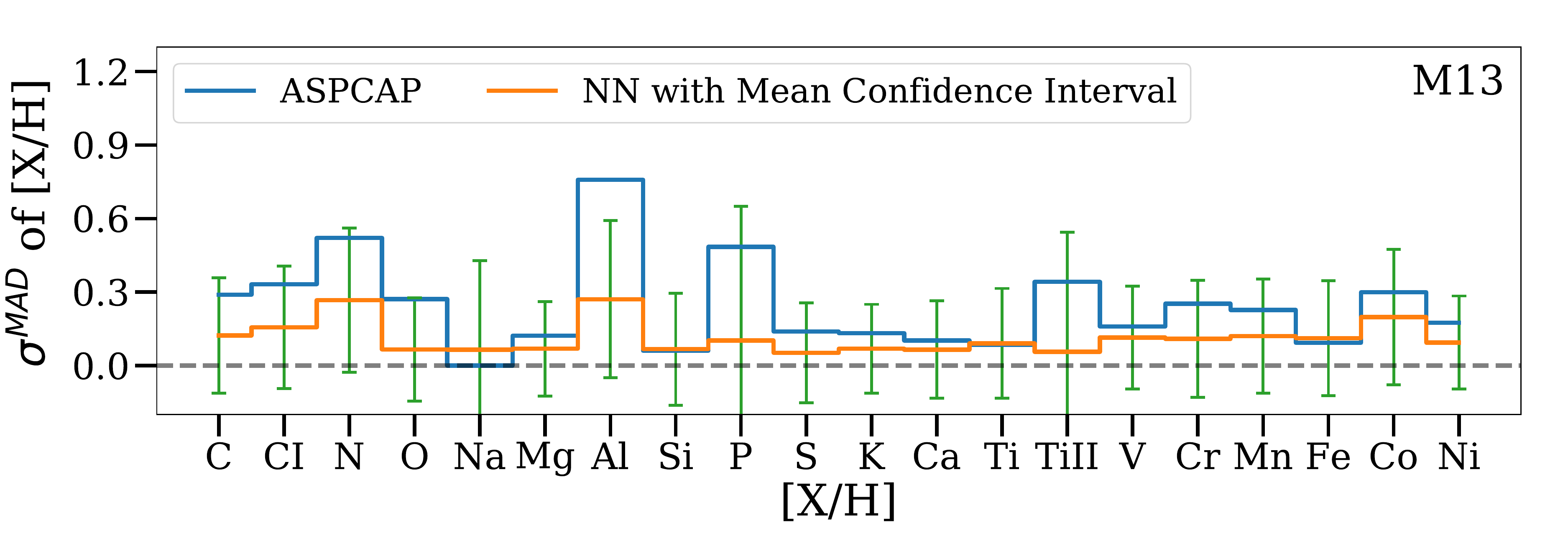}
\caption{Spread in the predicted NN and ASPCAP labels for stars in the globular cluster M13 (23 stars). As expected, the spread in the abundances of heavier elements is small, while that in light elements is larger. This is especially the case for Al, which has a $\approx0.5\,\mathrm{dex}$ spread in M13 \citep{2015AJ....149..153M} that is well recovered by the NN predictions.}
\label{figure:m13}
\centering
\end{figure}

\begin{figure*}
\centering
\includegraphics[width=\textwidth]{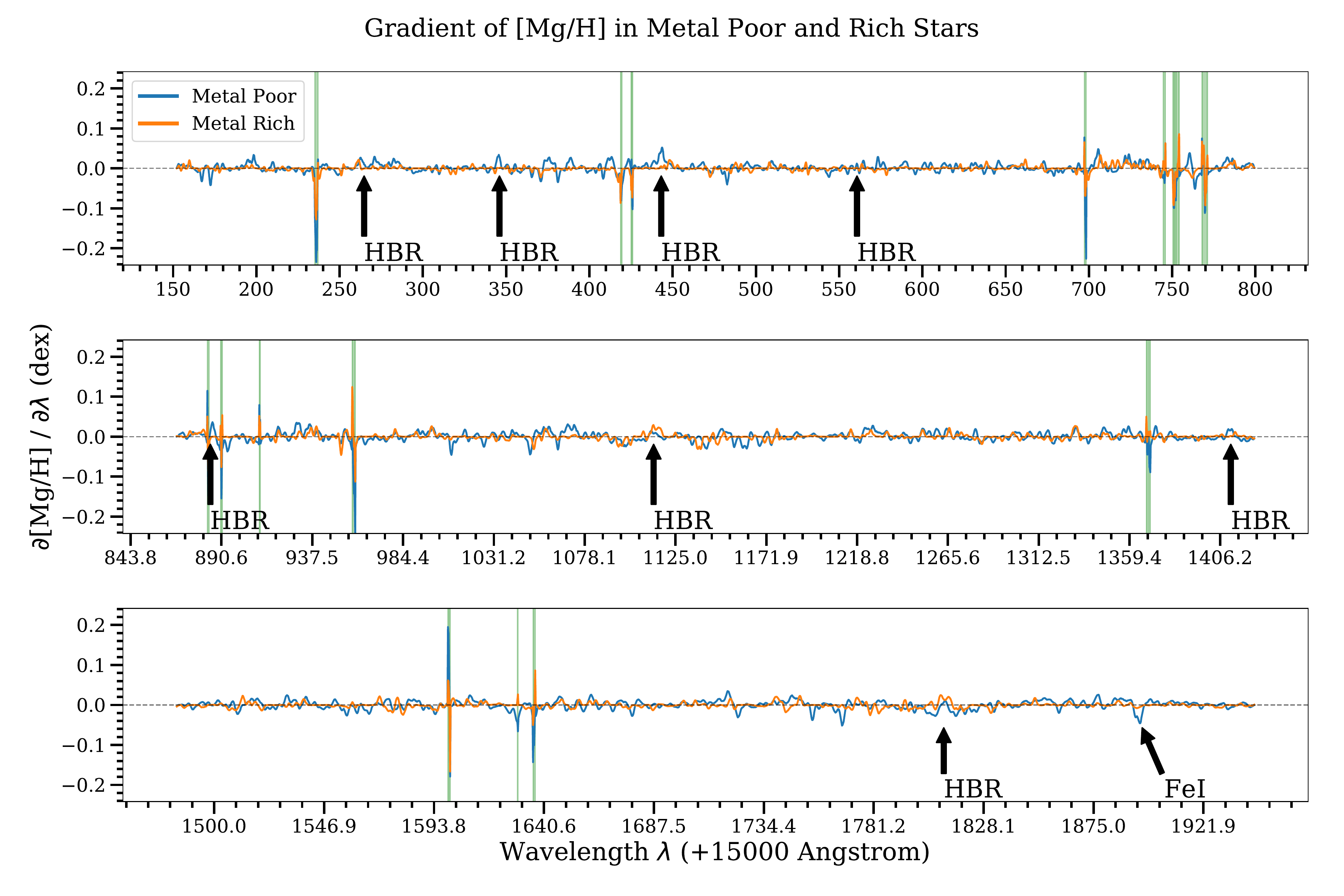}
\caption{The neural network's $\xh{Mg}$ sensitivity, that is mean $\frac{\partial\xh{Mg}}{\partial\lambda_j}$, for metal-poor ($\xh{Fe}<-1.5$; blue curve) and metal-rich stars ($\xh{Fe}>0.4$; orange curve).  The green regions show the ASPCAP \xh{Mg} windows. We also label the hydrogen lines and a strong FeI feature. Because of the way the network is structured, information about \xh{Mg} in the spectrum is mainly extracted in the ASPCAP windows, but especially for metal-poor stars the network also depends on features outside of the windows.}
\label{figure:jacobian}
\centering
\end{figure*}

Figure \ref{figure:snr_acc} shows the self-consistency of the neural network at low SNR in more detail. Here, the median absolute error is calculated between the predicted NN label for the individual exposure and the predicted NN label from the high-SNR combined counterpart. All abundances show significant errors at SNR$<30$. This is expected, because the noisy low-SNR spectra limit the ability of the neural network to get information from spectral features. However, the neural network still performs well at SNR$\approx 50$, with most abundances measured to a few hundredths of a dex. The performance of the neural network at SNR$\approx 50$ is better than that of the \texttt{Cannon 2} \citep{2016arXiv160303040C}, even though we only use regions of the spectra with known spectral features for the element of interest, while the \texttt{Cannon 2} uses the full spectrum for each element. The exact precision at SNR$\approx 50$ is shown in each panel, with the best performance of $0.014\,\mathrm{dex}$ for $\xh{Fe}$ and $\xh{Ca}$ to the worst performance of $0.068\,\mathrm{dex}$ for $\xh{P}$. Note that we measure \emph{each} element's abundance to better than $0.10\,\mathrm{dex}$, often considered the target uncertainty in large spectroscopic surveys, at SNR$\approx 50$. Of course, at lower metallicities the performance is worse, similar to what is seen in Figure \ref{figure:all_you_can_residue_highSNR}.

At high SNR, the residual in the predictions for most labels flattens out to $0.01 \text{ to } 0.02\,\mathrm{dex}$. This demonstrates that for most of the abundances, the neural network can reach a precision of $\approx 0.01\,\mathrm{dex}$ for high SNR spectra.

\subsection{Results on open and globular clusters}\label{subsec:clusters}

To further test the performance of the neural network's predictions for the individual elemental abundances, we apply it to open and globular clusters. Open clusters provide a good testbed for data-driven abundance analyses, because they are a chemically homogeneous population of stars, at least at the level of current abundance precision \citep[e.g.,][]{2006AJ....131..455D,2007AJ....133.1161D,2016ApJ...817...49B,2018ApJ...853..198N}. Ideally, the neural-network predictions for the abundances of stars in open clusters should exhibit a very small spread. Similarly, globular clusters should be homogeneous in all but their lightest elements. For the light elements we expect to see spreads and anti-correlations between pairs of elements (e.g., Mg and Al; \citealt{2015AJ....149..153M}). 

To perform these tests, we select stars from two open clusters that are well populated in the APOGEE database: M67 and NGC6819. We select members from the catalog provided by \citet{2013AJ....146..133M} and we exclude spectra with the \texttt{APOGEE\_STARFLAG} bitmask set. The resulting spreads in the abundances of each element in each cluster are shown in Figure \ref{figure:open_clusters}. The overall level of chemical homogeneity in these two clusters found by using the NN predictions is $0.030 \pm 0.029\,\mathrm{dex}$, obtained by calculating the mean abundance scatter and the mean uncertainty in the scatter. This is the same level of homogeneity as that reported by using the \texttt{Cannon 2} \citep{2018ApJ...853..198N} and \texttt{The Payne} \citep{2018arXiv180401530T} by benchmarking on open clusters.

\begin{figure*}
\centering
\includegraphics[width=\textwidth]{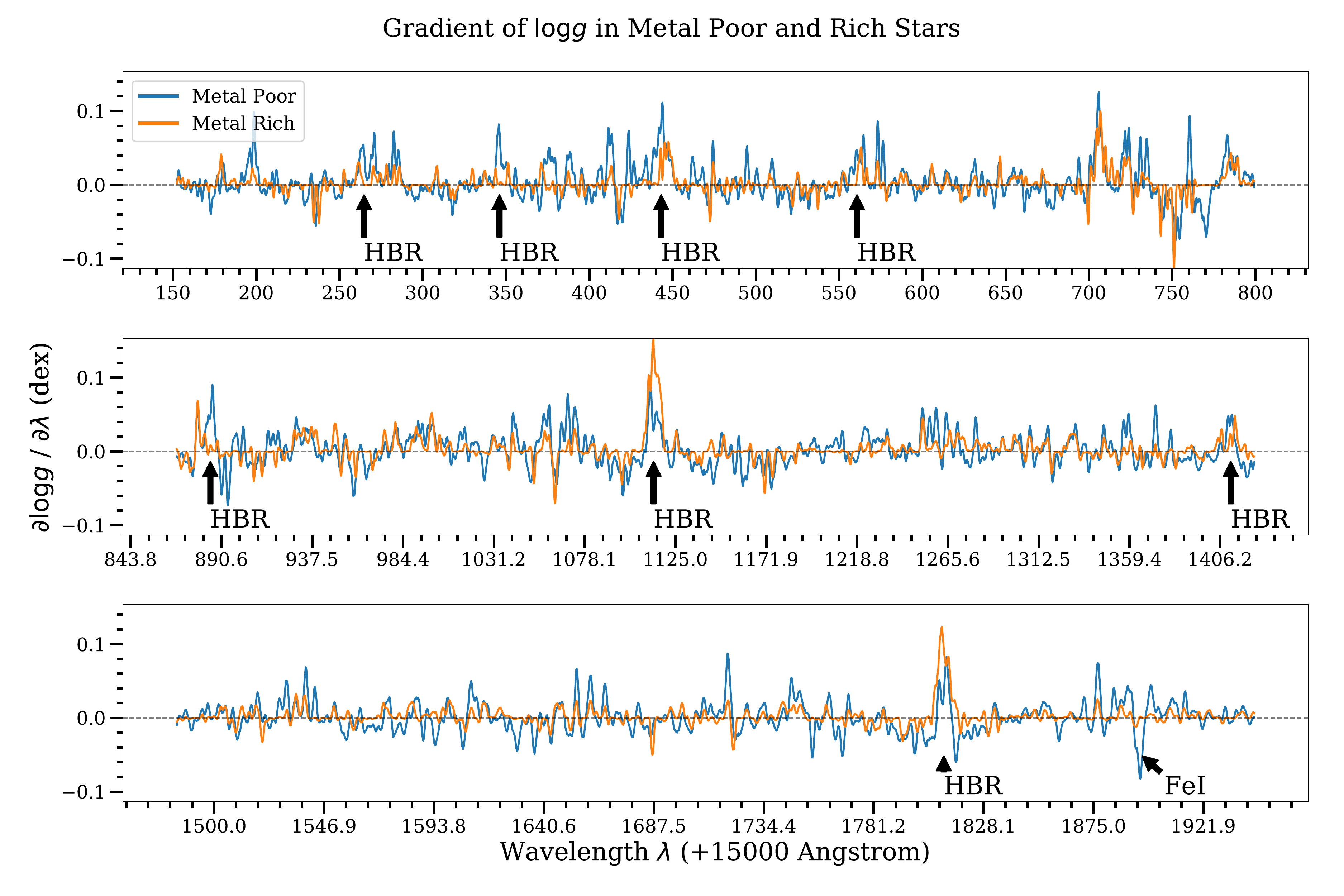}
\caption{The neural network's $\logg$ sensitivity, that is mean $\frac{\partial\logg}{\partial\lambda_j}$, for metal-poor ($\xh{Fe}<-1.5$; blue curve) and metal-rich stars ($\xh{Fe}>0.4$; orange curve). We also label the hydrogen lines and a strong FeI feature. Because of the way the network is structured, information about $\logg$ in the spectrum is mainly extracted in the hydrogen lines.}
\label{figure:jacobian_logg}
\centering
\end{figure*}

To test the neural network's performance at low metallicity, we use the globular cluster M13, which has many members in the APOGEE catalog. The spread in the abundances from the NN and from ASPCAP for M13 is displayed in Figure \ref{figure:m13}. This figure shows the expected behavior: the spread in the abundances of heavier elements is small, while that in the abundance of lighter elements is larger, especially for Al. A boutique analysis of the APOGEE spectra for stars in M13 by \citet{2015AJ....149..153M} showed that the spread in Al in M13 is particularly large: $\approx0.5\,\mathrm{dex}$. This is similar to the spread in the NN Al abundances in M13, while that in ASPCAP is significantly larger. In Figure \ref{figure:m13_al_mg} displayed in Section \ref{subsec:full_spec} below, we show the abundances of both Mg and Al for stars in M13 and these fall roughly along the expected sequence (which in M13 is almost vertical, that is, there is very little Mg spread in M13; \citealt{2015AJ....149..153M}).

\subsection{Sensitivity analysis}\label{subsec:jac}

In order to better understand what the neural network is doing, we can compute the sensitivity of each label to the input spectrum as this provides a glimpse into how the neural network makes its predictions and which regions in wavelength space are crucial for the neural network to predict each label. In mathematical terms, every neural network maps inputs $x$ to outputs $y$ in a differentiable manner (indeed, this differentiability is crucial in allowing the neural network to be optimized by gradient descent). In practice, neural-network frameworks allow this gradient to be computed analytically by making use of automatic differentiation. This procedure can be applied to compute the derivatives $\frac{\partial\text{Label}}{\partial\lambda}$ of each label with respect to every wavelength pixel $\lambda$ of the input spectrum. This derivative represents the sensitivity of the neural network to each pixel for every label. A negative $\frac{\partial\text{Label}_i}{\partial\lambda_j}$ indicates that if the flux at the $j^{th}$ wavelength bin $\lambda_j$ goes up, the value of the $i^{th}$ label decreases and vice versa for a positive value of $\frac{\partial\text{Label}_i}{\partial\lambda_j}$. 

An example of this type of sensitivity analysis is shown in Figure \ref{figure:jacobian}. This figure displays the derivative $\frac{\partial\xh{Mg}}{\partial\lambda_j}$, that is, the sensitivity of the neural network for the \xh{Mg} abundance. The derivative is averaged over two sets of stars in the high-SNR test set: all metal-poor stars with $\xh{Fe}<-1.5$ and all metal-rich stars with  $\xh{Fe}>0.4$. The green regions in this figure show the ASPCAP windows used to derive the \xh{Mg} abundance and the same windows that we use when making the \xh{Mg} prediction with the neural network. It is clear that for the metal-rich stars which have strong Mg features, the neural network mainly pays attention to the regions of the spectrum within the ASPCAP windows and only limited attention to the rest of the spectrum (recall that our neural-network architecture is such that a limited amount of information about the full spectrum can be used in the \xh{Mg} prediction, through the connection between the large neural network that predicts the $[\teff,\logg,\xh{Fe}]$ parameters and the mini-network that predicts \xh{Mg}). For metal-poor stars, the network still pays much attention to the region of the spectrum within the ASPCAP windows, but it also pays stronger attention to regions outside of the windows (and example is the strong FeI feature in the red part of the spectrum). This is because for metal-poor stars, the spectral Mg features are weaker, and so the neural network can improve its predictions by making use of a limited amount of information from the full spectrum. This shows that letting the neural network see the whole spectra is essential for it to make sensible predictions in extreme cases.

The behavior for other individual elements is similar to that for \xh{Mg} shown in Figure \ref{figure:jacobian}. For the \teff\ and \logg\ predictions, the neural network uses the entire spectral range. For \teff\ the network mainly gets information from a large number of spectral features. For \logg, we display the derivative  $\frac{\partial\logg}{\partial\lambda_j}$ in Figure \ref{figure:jacobian_logg}. While the derivative is non-zero over the full wavelength range, it is especially large near the hydrogen lines (the brackett series). This behaviour makes sense as the strong hydrogen lines are known to be strongly sensitive to \logg. Therefore, we see that even when the neural network is allowed to use the full spectrum, it uses a physically-plausible set of features in the spectrum to predict \logg.

\section{Variations}

\subsection{Training on the full, uncensored spectrum}\label{subsec:full_spec}

Before we settled on the \verb|ApogeeBCNNCensored()| NN architecture shown in Figure \ref{figure:nn_flow} that uses censored versions of the spectrum over the full wavelength range when determining the abundances of individual elements, we attempted using a simple multi-layered Bayesian convolutional NN with dropout. This network is available as \verb|ApogeeBCNN()| in the \texttt{astroNN} python package. Rather than splitting the label determination into a large NN to infer the main stellar parameters $[\teff,\logg,\xh{Fe}]$ and mini-networks to determine individual element abundances, this simple neural network is trained on the full spectra to infer all 22 parameters and abundances without any censorship. The performance of this \verb|ApogeeBCNN()| network on both the high-SNR and the individual-visits
test sets is similar to that of \verb|ApogeeBCNNCensored()| described in Sections \ref{subsec:highsnr} and Section \ref{subsec:lowsnr} for all labels. Thus, there is almost no loss in information in using only the censored spectra to determine individual abundances.

\begin{figure}
\centering
\includegraphics[width=0.5\textwidth]{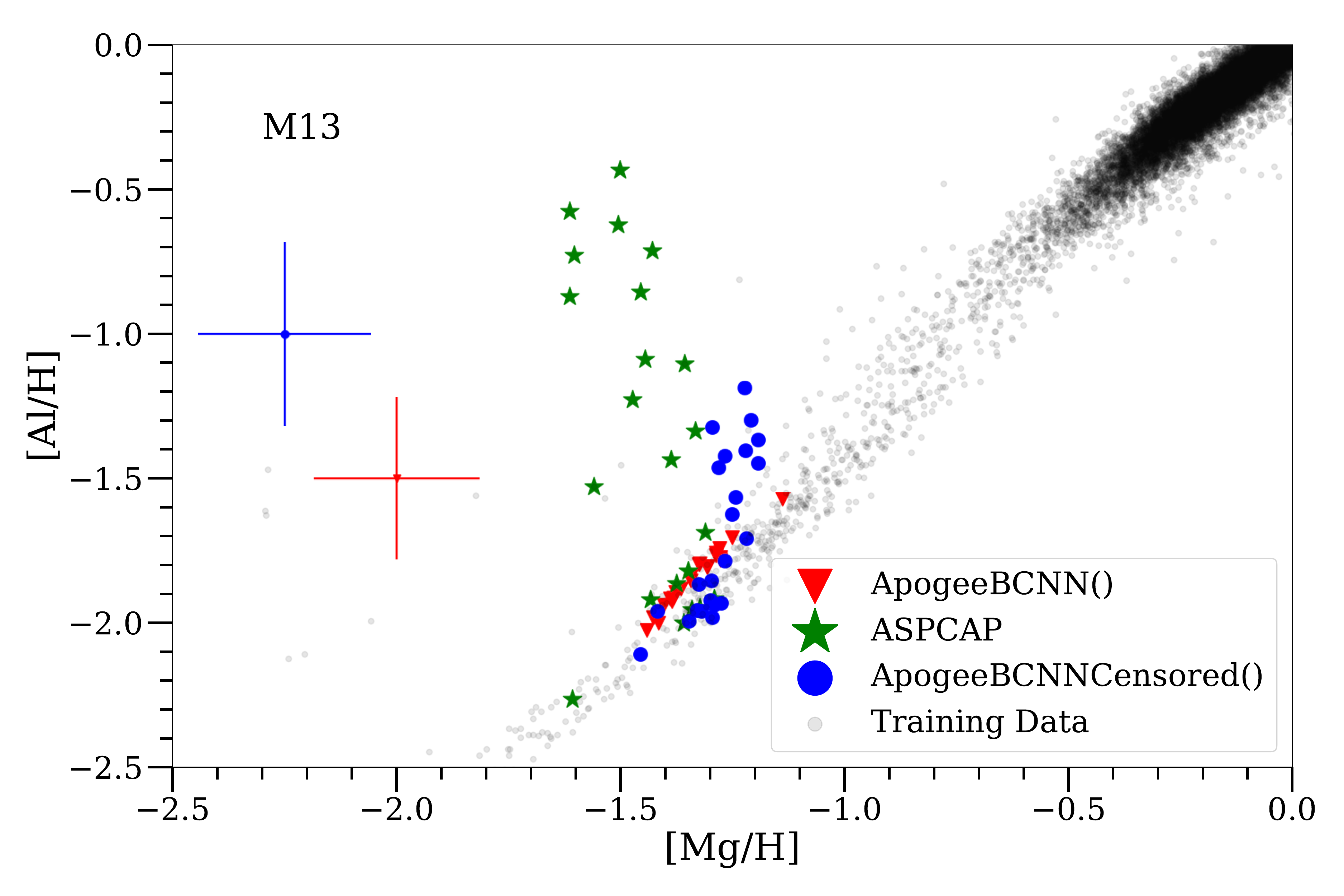}
\caption{\xh{Al} vs. \xh{Mg} as determined by the \texttt{ApogeeBCNNCensored()} NN, which only uses regions of the spectrum containing spectral features for each element to be determined, and by the \texttt{ApogeeBCNN()} NN, which uses the full spectrum for each element, for stars in the globular cluster M13. The black points show the distribution of \xh{Al} vs. \xh{Mg} in the training set while the blue and red point with errorbars show the median error with \texttt{ApogeeBCNNCensored()} and \texttt{ApogeeBCNN()} respectively. Because little information about \xh{Al} is available from regions containing Al features for M13 stars, the \texttt{ApogeeBCNN()} predictions of \xh{Al} follow the correlation with \xh{Mg} that is present in the data set. The \texttt{ApogeeBCNNCensored()} NN avoids this and displays the correct Al spread in M13 and show a better agreement with the analysis of \citet{2013AJ....146..133M} than ASPCAP does.}
\label{figure:m13_al_mg}
\centering
\end{figure}

However, the \verb|ApogeeBCNN()| network fails to perform well in regions of abundance space that are not well covered by the training set and where the intrinsic abundance trends are different from those for the majority of the sample. This is clearly seen when we apply the \verb|ApogeeBCNN()| network to the M13 globular cluster. As discussed in Section \ref{subsec:clusters} above, M13, like many globular clusters, has a wide spread in Al abundances and the Al abundances are anti-correlated with the Mg abundances (although for M13 the actual spread in Mg is very small; \citealt{2015AJ....149..153M}). In Figure \ref{figure:m13_al_mg}, we show the \xh{Al} vs. \xh{Mg} abundances for stars in M13 for the \verb|ApogeeBCNN()| and \verb|ApogeeBCNNCensored()| networks, as well as those for all stars in the training set. It is clear that for \verb|ApogeeBCNN()|, \xh{Al} is very strongly correlated with \xh{Mg} in M13. This likely results from the fact that the \xh{Al} abundance in M13 is difficult to measure (in large part because for the epoch at which most M13 stars were observed by APOGEE, one of the prominent Al lines in the $H$ band overlapped with a strong sky-emission line, rendering it unusable). In the absence of information on \xh{Al} from Al lines, \verb|ApogeeBCNN()| falls back on the correlation between \xh{Al} and \xh{Mg} seen in the training set and therefore, the \verb|ApogeeBCNN()| \xh{Al} and \xh{Mg} are almost entirely correlated. This happens because \verb|ApogeeBCNN()| does not know which features in the full spectrum belong to \xh{Al} and which to \xh{Mg}. 

Because the \verb|ApogeeBCNNCensored()| network uses only regions of the spectrum with Al features to determine \xh{Al}, it provides \xh{Al} measurements that are more in line with the results from \citet{2015AJ....149..153M}. As discussed above, the spread in \xh{Al} as determined by \verb|ApogeeBCNNCensored()| is about the same as that determined by \citet{2015AJ....149..153M}.

Despite the fact that the neural network trained on the full spectrum fails in certain regions of parameter space, we do allow a limited amount of information from the full spectrum to be used when determining the individual-element abundances. On the one hand these are the determinations of the overall stellar parameters $[\teff,\logg,\xh{Fe}]$, but we also include an additional, trainable two-neuron connection in the censored network as shown in Figure \ref{figure:nn_flow}. This allows the abundance prediction to depend on the full spectrum in a way that is not predetermined by us, but is learned from the training set. Such a connection makes physical sense, because the abundance of certain elements has a strong effect on the structure of the stellar photosphere, which in turn affects all parts of the spectrum. This is the case, for example, for carbon and oxygen in the cool stars observed by APOGEE \citep[e.g.,][]{2012AJ....144..120M}, but is also the case for other elements that are strong electron donors (e.g., Mg, Si), which therefore may also cause small effects through the spectral region. The two-neuron connection in \verb|ApogeeBCNNCensored()| allows such effects to be determined directly from the training data in a limited manner, without letting abundances of individual elements be determined entirely through correlations with other elements in the training data.

\subsection{Training with small data sets}

We have trained our neural network using high-SNR, high-resolution APOGEE spectra. Such high-quality data are expensive to obtain, because they require a large amount of telescope time. If we want to use a data-driven approach such as the neural network trained here to ``transfer'' labels from a high-resolution, high-SNR survey (e.g., APOGEE) to a low-resolution, low-SNR survey (e.g., LAMOST, as done using the \texttt{Cannon 2} by \citet{2017ApJ...836....5H}, we need to obtain a number of spectra for stars in common between the surveys, which takes away from the ability to observe new targets. Therefore, we investigate in this section whether we can train the neural network with smaller data sets and retain the good performance of the approach.

In traditional machine-learning techniques, the number of parameters is typically kept smaller than the size of the training data, because otherwise the machine-learning method will end up overfitting to the training data and therefore fail to generalize to the test data. However, a modern, medium-sized ANN often has billions of parameters that are optimized with less than a million training data. In our application, the network we use in Section \ref{sec:main_nn} has $\approx 3.5$ million trainable parameters, while the training set has $\approx 30,000$ objects. Therefore, the number of trainable parameters is more than a hundred times the number of training data. However, each spectrum consists of 7,514 flux values at different wavelengths, so the total number of training data points is $\approx 200$ million, more than the number of parameters. Therefore, we expect that we may be able to train the network with even fewer training objects.

The key to being able to train a large ANN with smaller training data sets lies in regularization \citep{2016arXiv161103530Z}. In our case, this regularization is provided by the dropout variational inference method used as a Bayesian approximation. Without this regularization, the network would simply memorize the results for the training set. The dropout procedure reduces the effective capacity of the network, by not allowing this type of memorization to work, and therefore helps the network to generalize to the test data.

We may ask whether we can reduce the number of training data. To do this, we train neural networks with exactly the same architecture and parameters as \texttt{ApogeeBCNNCensored()} but with training sets that are factors of $2^k$ smaller, from $2$ to $32$. The resulting bias and scatter of the residuals for the individual visit test set are displayed in Table \ref{table:smalldata_result} for the three main stellar parameters \teff, \logg, and \xh{Fe} (we do not show the results for the individual abundances, but these display similar trends). The bias in $\teff$ stays around $0\,\mathrm{K}$ for any size training set, similar to the $\teff$ bias shown in Table \ref{table:indi_result}. Compared to training on the whole training set, the network trained on 4,175 spectra (which is $12.5\%$ of the original training set) has larger scatter by $5$K, $0.012\,$dex and $0.003\,$dex in $\teff$, $\logg$, and $\xh{Fe}$ without introducing too much bias. Since the test set is mostly dominated by solar abundance stars, metrics in Table \ref{table:smalldata_result} are also mostly dominated by those stars. In regions of parameter space that are sparsely covered by the original training set, the performance becomes much worse with small training set. With even smaller training sets, the scatter across the whole parameters space as well as NN uncertainty increases significantly. Therefore, what is more important than the overall size of the training set is that it covers a wide range of possible parameter space.

Thus, we can obtain almost the same performance with a network trained on only a few thousand stars. Usage of the neural-network approach described in this paper for transferring APOGEE labels to surveys like LAMOST therefore look promising.

\begin{table}
  \centering
  \caption{Training on small data sets: median and \madstd\ of the \teff, \logg, \xh{Fe} residuals between the NN prediction and ASPCAP results for the individual test set when the NN is trained with a limited amount of data. Each label column shows median / \madstd.}
  \label{table:smalldata_result}
	\begin{tabular}{lccc} 
		\hline
        \# of objects \\ (7,514 pixels each) & \teff\ (k) & \logg\ (dex)  & \xh{Fe} (dex) \\
		\hline
        $33407$ & $-1\ / \ 24$ & $0.000\ / \ 0.046$ & $-0.004\ / \ 0.018$ \\
        $16703$ & $3\ / \ 24$ & $0.010\ / \ 0.055$ & $-0.005\ / \ 0.019$ \\
        $8351$ & $1\ / \ 27$ & $0.011\ / \ 0.059$ & $-0.006\ / \ 0.022$ \\
        $4175$ & $1\ / \ 28$ & $0.007\ / \ 0.058$ & $-0.004\ / \ 0.021$ \\
        $2087$ & $-4\ / \ 35$ & $0.013\ / \ 0.083$ & $-0.014\ / \ 0.029$ \\
        $1043$ & $9\ / \ 53$ & $0.03\ / \ 0.16$ & $-0.012\ / \ 0.039$ \\
		\hline
	\end{tabular}
\end{table}

\subsection{Importance of continuum normalization}

\begin{table}
  \centering
  \caption{Neural-network prediction results on the high-SNR test set from comparing to ASPCAP, when using ASPCAP's procedure for continuum-normalization on the training and testing spectra.}
    \label{table:aspcapnorm_result}
	\begin{tabular}{lrr}
		\hline
        Label & Median of residual & \madstd\ of residual\\
		\hline
        $\teff$ & $-22 \text{ K}$  & $31 \text{ K}$ \\
        $\logg$ & $0.010 \text{ dex}$  & $0.050 \text{ dex}$ \\
        $\xh{C}$ & $-0.003 \text{ dex}$ & $0.051 \text{ dex}$ \\
        $\xh{CI}$ & $0.010 \text{ dex}$  & $0.056 \text{ dex}$ \\
        $\xh{N}$ & $0.008 \text{ dex}$  & $0.064 \text{ dex}$ \\
        $\xh{O}$ & $-0.017 \text{ dex}$  & $0.047 \text{ dex}$ \\
        $\xh{Na}$ & $-0.01 \text{ dex}$  & $0.13 \text{ dex}$ \\
        $\xh{Mg}$ & $-0.000 \text{ dex}$  & $0.025 \text{ dex}$ \\
        $\xh{Al}$ & $-0.040 \text{ dex}$ & $0.070 \text{ dex}$ \\
        $\xh{Si}$ & $-0.002 \text{ dex}$  & $0.029 \text{ dex}$ \\
        $\xh{P}$ & $-0.02 \text{ dex}$  & $0.11 \text{ dex}$ \\
        $\xh{S}$ & $0.011 \text{ dex}$  & $0.060 \text{ dex}$ \\
        $\xh{K}$ & $-0.010 \text{ dex}$  & $0.046 \text{ dex}$ \\
        $\xh{Ca}$ & $-0.014 \text{ dex}$  & $0.030 \text{ dex}$ \\
        $\xh{Ti}$ & $-0.024 \text{ dex}$  & $0.050 \text{ dex}$ \\
        $\xh{TiII}$ & $-0.04 \text{ dex}$  & $0.11 \text{ dex}$ \\
        $\xh{V}$ & $-0.007 \text{ dex}$  & $0.099 \text{ dex}$ \\
        $\xh{Cr}$ & $0.000 \text{ dex}$  & $0.048 \text{ dex}$ \\
        $\xh{Mn}$ & $-0.016 \text{ dex}$  & $0.038 \text{ dex}$ \\
        $\xh{Fe}$ & $-0.003\text{ dex}$  & $0.018 \text{ dex}$ \\
        $\xh{Co}$ & $-0.02 \text{ dex}$  & $0.14 \text{ dex}$ \\
        $\xh{Ni}$ & $0.003 \text{ dex}$  & $0.030 \text{ dex}$ \\
		\hline
	\end{tabular}
\end{table}

\begin{figure*}
\centering
\includegraphics[width=\textwidth]{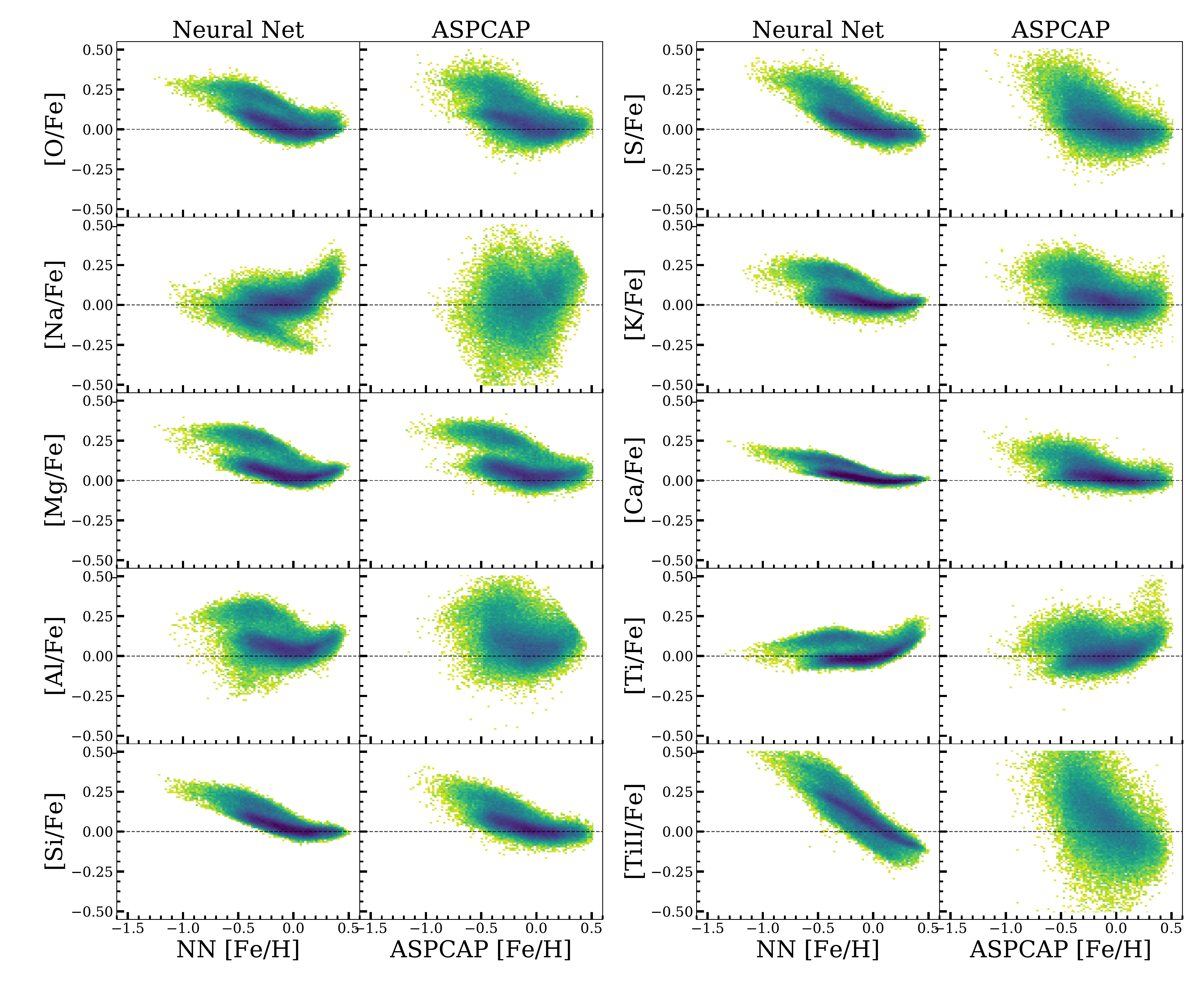}
\caption{Neural network and ASPCAP predictions on 157,598 APOGEE DR14 stars after a cut on neural network \logg\ uncertainty $<0.2\,\mathrm{dex}$ as a means to filter out (a) main-sequence stars (as discussed in the right panel of Figure~\ref{figure:logg_teff_isochrones} ) and (b) problematic spectra that  result in a high uncertainty in labels including \logg. This figure shows all $\alpha$ and odd-Z light elements Na, Al, and K among our 20 labels prediction. For most abundances, the NN abundances display less scatter and a clearer high/low alpha sequence than ASPCAP.}
\label{figure:xfe_alpha}
\centering
\end{figure*}

\begin{figure*}
\centering
\includegraphics[width=\textwidth]{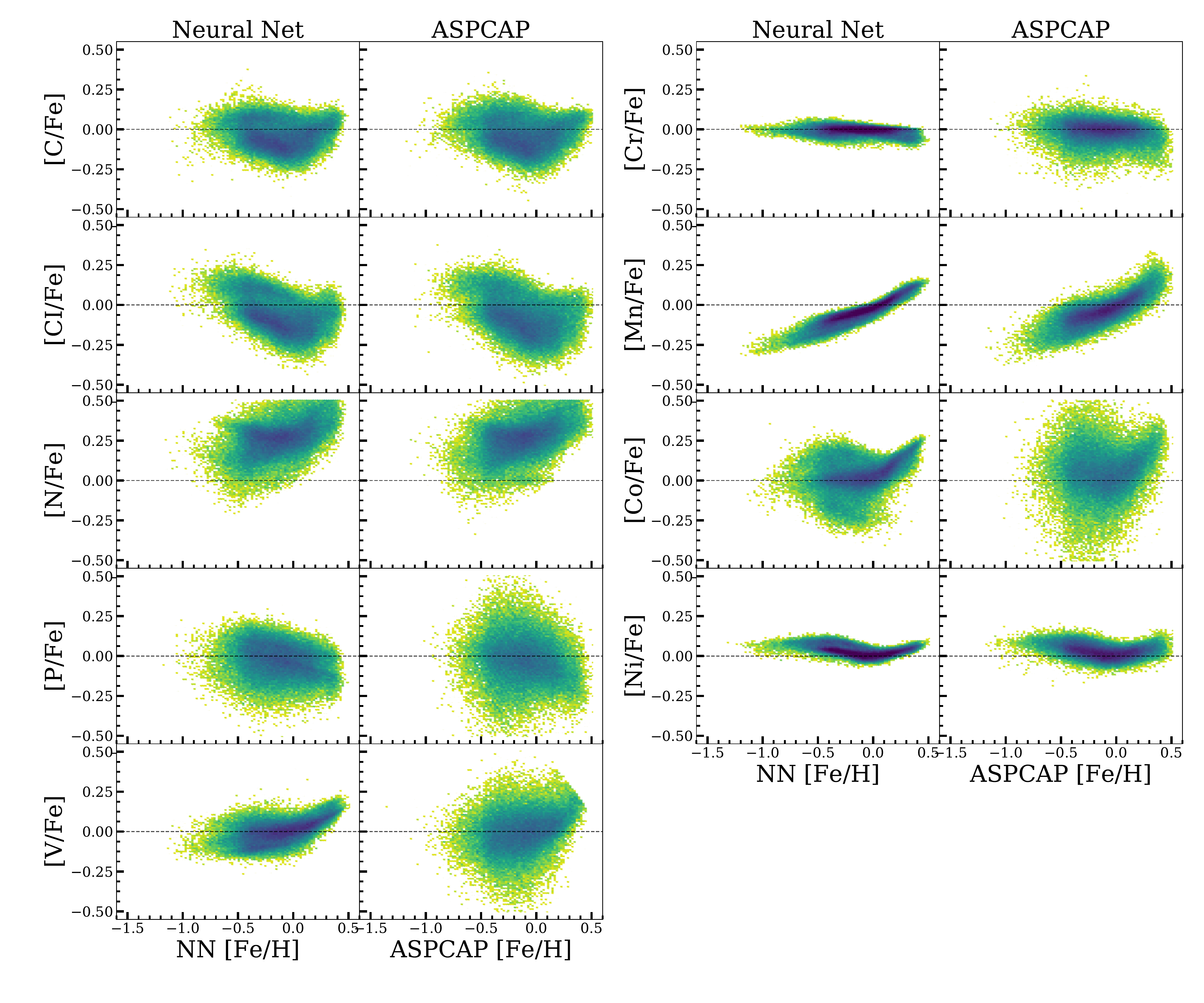}
\caption{Like Figure \ref{figure:xfe_alpha}, but for the remaining elements. For abundances like \xfe{Cr} and \xfe{Ni}, the NN abundances show much tighter scatter at \xfe{X}=0 as expected.}
\label{figure:xfe_not_alpha}	
\centering
\end{figure*}

As described in Section \ref{subsec:reduction}, we have chosen to employ a custom continuum normalization procedure that uses a set of pixels assumed to represent the (pseudo-)continuum on both individual exposures and on the combined spectra, instead of using ASPCAP's pseudo-continuum normalized spectra, which simply fit a polynomial to each detector's spectrum. Both of these procedures are expected to be independent of SNR, but the procedure we use produces continuum-normalized spectra that are more similar to those used in traditional spectroscopic analyses. In this section, we test what happens when we use ASPCAP's procedure instead.

Table \ref{table:aspcapnorm_result} shows the result of a \verb|ApogeeBCNNCensored()| neural network trained on the same data but with spectra that are continuum normalized using ASPCAP's procedure (we simply use the continuum-normalized spectra provided in the APOGEE data release for this, rather than attempting to reproduce ASPCAP's normalization). We then similarly test on the same testing data as in the high-SNR test set, but with spectra normalized using ASPCAP's procedure. Comparing the results in Table \ref{table:aspcapnorm_result} with those in Table \ref{table:highsnr_result}, we see that the resulting biases are similar, but the scatter in some of the abundance labels (e.g., \xh{C} and \xh{N}) is considerably larger. This test therefore demonstrates that the continuum normalization that we use does improve the neural network's performance, so there is use in attempting to obtain a (pseudo-)continuum that is close to the true continuum.

\section{Abundance distributions in the Milky Way}

To further illustrate the power of the NN approach, we show the \xfe{X} vs. \xh{Fe} distributions of stars for all elements measured by ASPCAP and compare it to the ASPCAP results for the same stars. We only use a single cut to the full APOGEE DR14 catalog of spectra: we remove all stars with \logg\ uncertainty larger than 0.2 dex. This cut removes any problematic spectra which result in large label uncertainties (traced by the \logg\ uncertainty) and also cuts out essentially all dwarfs; because the NN \logg\ results for dwarfs are based on extrapolation---the training set contains no \logg\ for dwarfs---the actual NN \logg\ values are very wrong, so it is essential to cut on their uncertainty instead (see Figure \ref{figure:logg_teff_isochrones}). After this cut, we are left with a sample of 157,598 stars.

Figure \ref{figure:xfe_alpha} displays the abundance distribution for $\alpha$ and odd-Z light elements Na, Al, and K  for these stars, both for the NN and for ASPCAP. Figure \ref{figure:xfe_not_alpha} shows the distribution of the remaining elements. As above, we show results for CI and TiII separately, as these are measured separately by ASPCAP. It is clear that the NN abundance patterns are tighter than those obtained from ASPCAP. This is especially clear for the $\alpha$ elements (O, Mg, Si, S, Ca, Ti), which has a distinct bimodal structure at intermediate metallicities ($\xh{Fe} \approx -0.2$ to $-0.5$) in \emph{all} $\alpha$ elements. The low-$\alpha$ sequence in all of these distributions is also significantly tighter than in the ASPCAP results. We also see that Al and K essentially behave as $\alpha$ elements, both displaying a clear bimodal structure with the same pattern as the $\alpha$ elements.

\section{Discussion}

\subsection{Performance}

Deep learning is a promising tool for the big-data era in astronomy. Besides the better precision and accuracy of the neural network compared to other approaches to spectroscopic analysis, the development of modern hardware-accelerated deep-learning technology means that our analysis is also much faster than traditional and other data-driven approaches. The neural network described in this work determines 22 labels from $\approx 100,000$ APOGEE spectra consisting of $\approx 7,500$ wavelength pixels each, while doing $100$ Monte Carlo dropout runs---equivalent to determining 22 parameters from 10,000,000 APOGEE spectra without dropout---in $\approx 300$ seconds or $\approx3\,\mathrm{ms}$ per star ($\approx30\,\mu\mathrm{s}$ per star without dropout). This performance is obtained on a Nvidia consumer graphics processor with $4,375$ GFLOPS at single precision, while the training steps with 40 epochs and a batch size of 64 takes $\approx 700$ seconds to complete in total\footnote{\textit{Batch size} refers to the number of training examples in one forward/backward pass and \textit{epoch} refers one forward pass and one backward pass of all the training examples. For example, with 6400 training data and a batch size of 64 with 5 epochs, each epoch will perform 100 gradient updates, each gradient is an average of 64 training data without replacement, and do it 5 times.}. The performance of neural networks is expected to get much better with upcoming graphics processors specializing in deep learning, such as Tensor Processing Units (TPUs) in upcoming consumer Nvidia GPUs \citep{2018arXiv180406826J}. 

This fast performance offers great advantages in the era of large spectroscopic surveys. For example, it means that computational needs are far lower when only a small, high-SNR subset of the data (the training set) is analyzed with traditional, slow tools, while the majority of the data set can be processed much faster with the neural-network approach. Currently, the APOGEE ASPCAP pipeline requires a large cluster both to produce the library of synthetic spectra used in the ASPCAP fitting and for performing the fits themselves. Our fast framework allows for a much faster development cycle. This is the case in the narrow sense of providing the opportunity for fast prototyping and exploration of new neural-network model architectures. But when coupled with development of the tools to produce the small training set, this is also the case in the broader sense of seeing how changes to the input physics used in producing the training set affect the larger test sample.

Our extremely fast analysis coupled with the fact that the dropout procedure produces realistic uncertainties also opens up the possibility of real-time analysis of whether high enough SNR is obtained for a given level of abundance uncertainties. Stellar spectra could be analyzed on-the-fly as they are being collected and integration stopped when an acceptable uncertainty in the abundances is reached. With future, large fibre-positioner systems this could be used to run efficient, large spectroscopic surveys. Because our approach automatically assigns large uncertainties for objects far outside the training-set boundaries, such objects, which are likely to be of interest, would automatically obtain high SNR spectra in this approach.

\subsection{Comparison to other data-driven approaches to spectral analysis} \label{subsec:assumption}

Another data-driven approach to high-resolution analysis is provided by the \texttt{Cannon} \citep{2016arXiv160303040C}, which we have already discussed in the text above. The \texttt{Cannon}'s approach is fundamentally different from ours, in that the \texttt{Cannon} builds a data-driven model of the spectra as a function of stellar labels that is then used to fit labels to observed spectra, while our approach directly determines the mapping from spectra to labels. We therefore make slightly different assumptions about the relation between spectra and labels. To be clear, we list our methods and assumptions (or lack thereof) and how they compare to those made by the \texttt{Cannon}:

\begin{itemize}
  \item NNs in this work map spectra directly to labels in a single step (and a sequence of these single steps to determine the uncertainty using dropout). The \texttt{Cannon 2} is a generative model that generates realistic spectra from labels and then matches spectra by $\chi^2$ minimization to find abundances.
  \item We assume that the value of the labels is a continuous, smooth function of the flux. Thus, we assume that similar spectra have similar labels. This assumption is shared by the \texttt{Cannon 2} and essentially by all approaches to spectroscopic fitting.
  \item Despite that, we do not require that spectra with similar labels are similar. This limits our ability to generate a new spectrum for certain set of labels and we cannot generate a set of  $\Delta$flux for a given spectrum that changes only a single label, while keeping the others constant. The \texttt{Cannon 2} assumes that spectra with similar labels are similar.
  \item The flexibility of the NNs mean that it only performs well on what it is trained and the ability to extrapolate is limited (but note that when extrapolation occurs, the returned uncertainties are very large). This is similar to the \texttt{Cannon 2}, although their simpler model may approximate physical models better and thus have better performance when extrapolating.
\end{itemize}

Our NN approach has some disadvantages with respect to forward-modeling approaches like the \texttt{Cannon}:\\
\emph{Interpretability:} Because the \texttt{Cannon} builds a forward model of the spectra it allows one to generate spectra from the model for a given set of labels. This can be used to inspect the internal functioning of the model and one can ask questions such as: does the behaviour when changing \xh{Al} while keeping all other parameters fixed make sense? Because of the reasons given above, our NN cannot answer such questions and therefore does not allow inspection of the model in this sense. It is possible to compute derivatives of each label value with respect to the input spectra (e.g., see Figures \ref{figure:jacobian} and \ref{figure:jacobian_logg}), but as these derivatives do not keep other labels fixed, they are less meaningful.\\
\emph{Handling of uncertainties in the input:} Our approach ignores uncertainties in the spectra. An approach that forward models the spectra can fit in both the training and test step while taking the flux uncertainties into account. Nevertheless, we have shown that even though we train on high SNR spectra, the NN performs well even at very low SNR. Part of the reason for this is that the flux uncertainties are relatively constant with wavelength. This is not the case for the label uncertainties in the training set (see above) and because it is far easier to only take one of input or output uncertainties into account in any machine-learning technique, it appears more important for the analysis of stellar spectra to take the label uncertainty into account.

On the other hand, the advantages of the NN approach over an approach like the \texttt{Cannon} are manifold:\\
\emph{Speed:} Because our method learns a mapping from spectra to labels and does not require any fitting when making predictions, it is much faster. As discussed above, we can determine labels for $\approx 100,000$ APOGEE stars in about three seconds on a single $\approx\$500$ GPU (and in five minutes if we also want the uncertainty), while running the \texttt{Cannon 2} on the same number of spectra takes about 20 minutes on a small cluster \citep{2016arXiv160303040C}.\\
\emph{Realistic uncertainties:} The \texttt{Cannon 2} can obtain uncertainty estimates from its $\chi^2$ fitting to the spectra and their uncertainties, but these uncertainties are typically much smaller than the scatter in the results obtained from cross-validation, demonstrating that the uncertainties are underpredicted. Our approach returns uncertainties that, at least within the well-populated regions of the training set, are approximately equal to the scatter from cross-validation or equal to the scatter in open clusters. Thus, our uncertainties can be used in practice to determine whether observed scatter is real or due to noise in the data.\\
\emph{Extrapolation warnings:} Data-driven approaches only perform well within the training set and do not perform well when extrapolating, especially approaches as flexible as our NN approach. Data-driven approaches therefore typically need a way to determine first whether or not a test object is within the bounds of the training set. For high-dimensional input and label spaces this boundary can be a complicated, high-dimensional surface that is difficult to determine. In tests, this issue is typically ignored, as most test sets are chosen to represent the training set (e.g., in the standard random partition of the full data set into a training and test set). As we demonstrated above, the uncertainties returned by our NN are very large whenever a spectrum or label outside of the training set is analyzed (e.g., the main-sequence stars' \logg\ in Figure \ref{figure:logg_teff_isochrones}). These large uncertainties can therefore be used as a warning flag for spectra or labels outside of the training set and our method thus provides an automatic check for the training-set boundaries.\\
\emph{Training on noisy and incomplete data:} Because we take the uncertainties on the training labels into account and can also use training stars for which not all labels are measured, we can train on an imperfect training set. This is important, because for various reasons many stars in the training set, even at high SNR, do not have complete and high-precision set of labels. The \texttt{Cannon 2} in its current implementation does not take uncertainties in the training labels into account and can only train on training objects with a complete set of labels.

The advantage of all data-driven approaches is that they allow properties such as stellar mass, age, or luminosity to be determined directly from stellar spectra. This is difficult to do with theoretical stellar models, because how these properties affect the spectra is not well known. For example, a common way to infer luminosity is by using isochrones based on stellar evolutionary models and stellar parameters and abundances measured from stellar spectra \citep{2016AA...585A..42S}, but stellar isochrones are not well calibrated for all types of stars and stages of stellar evolution. Spectra may contain direct indicators of mass \citep[e.g.,][]{2016ApJ...823..114N}, age, or luminosity that can be extracted by training on data sets for which these are known quantities (e.g., masses from asteroseismology, e.g., APOKASC: \citealt{2014ApJS..215...19P}; or luminosities from \emph{Gaia}).

\subsection{Comparison to neural-network approaches to spectral analysis based on synthetic spectra}\label{subsec:comparesynth}

The training set potentially does not represent all stars in the test set and, as discussed above, NNs trained on an unrepresentative training set will perform poorly for stars outside of the training set. Moreover, abundances determined by ASPCAP contain systematic errors that propagate to the NN during training and we cannot easily quantify these errors or correct them. These systematic errors result from the quality of ASPCAP synthetic grid and errors introduced by the data reduction and calibration steps in ASPCAP. One solution is using the techniques used by \texttt{StarNet} or \texttt{The Payne} \citep{2018arXiv180401530T} by training on theoretical synthetic spectra, because theoretical spectra with different combinations of abundances can be generated and we can choose these combinations to represent the likely set of abundances in the test set. In terms of performance, \texttt{StarNet} uses neural networks similar to those used here and, when implemented on a GPU, could be as fast to train and evaluate as our method. \texttt{The Payne}, however, exhibits far slower performance because it is a forward model of the spectra given the labels similar to the \texttt{Cannon}, but with higher complexity. Training \texttt{The Payne} takes about 5 CPU minutes per wavelength pixel or 26 days for all pixels, compared to our 10 minutes. Evaluating for a test spectrum takes about 1 CPU s per spectrum or about 28 hours for $100,000$ APOGEE stars, versus our 3 seconds for $100,000$ APOGEE stars (300 s including uncertainties).

Training on synthetic spectra can lead to bad performance when the theoretical spectra do not match the observed spectra well. This can be due to calibration, uncertainties, continuum-normalization issues, or unmodeled physics in the spectral synthesis. For example, without correcting systematic differences from the observed spectra before training, \texttt{StarNet} reports that their neural network trained on synthetic spectra does not perform well on observed spectra due to differences in the feature distributions between theoretical and real spectra. Of course, because data-driven approaches rely on a training set that is typically analyzed using synthetic spectra, they suffer from limitations in the input physics as well. However, through clever use of star clusters, binaries, and similar systems it may be possible to create training sets that are less sensitive to the theoretical modeling of stellar spectra. This is because if we can assume that these systems are chemically homogeneous, we can transfer the chemical labels from well-understood stellar types onto those of poorly-understood stellar types, thus creating an empirical training set for the poorly-understood types.

Another goal described in \texttt{Cannon 2} and the version of \texttt{StarNet} which trained on real world spectra is to identify potential unknown abundance lines. The use of a censored neural network in this work will render this practically impossible, because we limited the attention of the network to known spectral lines. However, even for a network trained on the full spectrum or on synthetic spectra, it will be questionable whether a previously unknown spectral feature that is identified really belongs to a certain element.

\section{Conclusions}

Large spectroscopic surveys are now routinely obtaining high-resolution spectra for hundreds of thousands of stars and upcoming surveys such as WEAVE \citep{2014SPIE.9147E..0LD}, SDSS-V \citep{2017arXiv171103234K}, 4MOST \citep{2012SPIE.8446E..0TD}, and MSE \citep{2016arXiv160600043M} will soon provide such spectra for millions of stars. Traditionally, such spectra are analyzed one-by-one with procedures that are largely manual, which is clearly impractical for large data sets. Fitting spectra with synthetic libraries (e.g., ASPCAP) is the currently favored approach, but this method is slow and current implementations do not deal well with low SNR spectra. As an alternative approach, we have presented neural network models for spectroscopic analysis to infer stellar parameters and chemical abundances with associated uncertainty using Bayesian ANNs with dropout variational inference. We implemented this method in a general, open-source python framework for ANNs called \texttt{astroNN} (see Appendix \ref{appendix:graph:astroNN}). We also release a catalog of abundances for the APOGEE DR14 data set determined by our \verb|ApogeeBCNNCensored()| network (see link in the Introduction).

Our neural-network method has various special ingredients beyond what would be used in a standard deep learning application. We (a) present a robust objective function for the neural network to learn from incomplete data while taking uncertainty in the training labels into account, (b) use a Bayesian neural network with dropout variational inference with this objective function to estimate the uncertainties on the labels determined by the neural network, and (c) combine a large NN to obtain the overall stellar parameters $[\teff,\logg,\xh{Fe}]$ with mini-networks to determine individual elemental abundances that use versions of the full spectrum censored to only include regions with spectral lines for a given element. With this approach, we simultaneously determine 22 stellar and elemental abundance labels accurately and precisely for both high and low SNR spectra. We implemented the method on a GPU using standard tools, which allows speed-ups of more than an order of magnitude and we make these tools easily accessible (see Appendix \ref{subsec:fastMC}). Our method is extremely fast, allowing stellar parameters and abundances for the entire APOGEE database to be determined in about ten minutes on a single GPU. 

We performed detailed tests of our method by comparing to results from the standard APOGEE ASPCAP pipeline and by comparing results on high-SNR, combined spectra to those from low-SNR, individual exposures. At high SNR, we obtain abundances precise to $\approx 0.01$ to $0.02\,\mathrm{dex}$ and even at low-SNR (SNR$\approx 50$) we get precisions of $\approx 0.02$ to $0.03\,\mathrm{dex}$ for most elements. These precisions are confirmed by looking at the scatter in the abundances within open clusters, which we find to be $0.03\pm0.03\,\mathrm{dex}$. We also recovered the expected abundance trends in globular clusters, but found that the censoring in the network is crucial to obtain this, because training on the full spectrum for all elements causes the NN to depend too strongly on correlations between elements within the training set and therefore to fail when these correlations are absent, like in globular clusters. We also demonstrated that a large neural network can work well with a limited amount of training data, finding barely degraded performance for training sets that only consists of thousands of spectra.

The speed and flexibility of neural networks mean that they are a highly useful tool for spectroscopic data analysis. They allow the results obtained from a detailed analysis of a small calibration set to be transfered to an entire large data set of millions of spectra in a matter of minutes and can thus be a great aid in the development of the next generation of spectroscopy tools. They could also be of use in earlier stages of the data processing, for example, for homogenizing spectra taken with different instruments (e.g., the northern and southern APOGEE spectrographs) or for removing instrumental systematics such as persistence \citep[e.g.,][]{2017MNRAS.470.4782J}. Such applications would be easy to pursue with the \texttt{astroNN} software package described in the appendix.

\section*{Acknowledgements}

It is our pleasure to thank Kim Venn and other members of the \texttt{StarNet} group for valuable feedback and for releasing their code publicly. We also thank Natalie Price-Jones for help with the APOGEE data. HL and JB received support from the Natural Sciences and
Engineering Research Council of Canada (NSERC; funding reference number RGPIN-2015-05235) and from an Ontario Early Researcher Award (ER16-12-061). JB also received partial
support from an Alfred P. Sloan Fellowship.

Funding for the Sloan Digital Sky Survey IV has been
provided by the Alfred P. Sloan Foundation, the U.S. Department
of Energy Office of Science, and the Participating
Institutions. SDSS-IV acknowledges support and resources
from the Center for High-Performance Computing at the
University of Utah. The SDSS web site is www.sdss.org.








\appendix
\onecolumn
\section{\texorpdfstring{\lowercase{astro}}NNN: a Python library for deep learning in astronomy}\label{appendix:graph:astroNN}

Data-driven tools are becoming more and more popular in astronomy and deep learning in particular is becoming a popular technique to deal with big data sets. When we started on the research for this paper, we realized that a python package that (a) has a focus on deep learning, (b) contains tools and data sets that are relevant to astronomers, (c) is easy to use and well tested against errors, and (d) acts as a platform to share astronomy-oriented neural networks is needed to advance research in this area.

\texttt{astroNN} is a python package for deep learning in astronomy designed to fulfill the points stated above. \texttt{astroNN}  relies heavily on \texttt{Tensorflow} \citep{2016arXiv160304467A}, a standard deep learning python library, and is easy to setup on all common platforms. \texttt{astroNN} employs custom methods for storing and sharing models: models are saved in a folder and consist of the neural network itself in \texttt{HDF5} format, the training history in \texttt{CSV} format, model parameters and the names of output neurons as well as normalization parameters in \texttt{JSON} format. Users simply have to give the reduced data as \texttt{numpy} array(s) to \texttt{astroNN} and get outputs back as \texttt{numpy} array(s); normalization of the inputs and outputs is handled internally.

At this point, \texttt{astroNN} as used in this paper consists of about 8,300 lines of code in the modules, 1,200 line of test code, and 3,200 lines of documentation. The test suite covers $>90\%$ of the neural network-related components of the code. Version control, continuous integration, test statistics, and documentation hosting is provided by \texttt{GitHub}, \texttt{Travis-CI}, \texttt{Coveralls} and \texttt{Read The Docs}, respectively. \texttt{astroNN}is available at \url{https://github.com/henrysky/astroNN} with extensive documentation at \url{http://astroNN.readthedocs.io}.

\subsection{Modules}

The main structure of \texttt{astroNN} is as follows: 

\begin{itemize}
  \item Three data processing modules which provide basic data reduction and processing for different surveys. \texttt{Astropy} \citep{2018arXiv180102634T} is used in these modules to provide unit conversions and \texttt{FITS} file reading functionality.
    \begin{itemize}
      \item \texttt{astroNN.apogee} $-$ Provides basic functionality to read APOGEE DR13/14 data sets and processing of the spectra.
      \item \texttt{astroNN.gaia} $-$ Provides basic functionality to read \emph{Gaia} DR1/2 data sets and astrometry-related conversion tools.
      \item \texttt{astroNN.lamost} $-$ Provides basic functionality to read LAMOST data sets and spectra processing.
    \end{itemize}
    \item Two neural-network related modules which provide different model architectures and infrastructure. All functions are compatible with \texttt{Tensorflow} or with \texttt{keras} with the \texttt{Tensorflow} backend \citep{keras2015}.
    \begin{itemize}
      \item \texttt{astroNN.models} $-$ Neural network architectures and neural network classes are defined in this module. Currently, this includes a convolutional neural network, a Bayesian neural network with dropout variational inference, as well as a variational auto-encoder for unsupervised spectra analysis. Note that the latter is still under development.
      \item \texttt{astroNN.nn} $-$ Deep-learning infrastructure such as robust objective functions, objective functions to deal with incomplete data, and customized layers such as Monte Carlo Dropout layer, Gradient Stopping layer or Boolean Masking layer that are not available in standard tools.
	\end{itemize}
    \item One data sets module which provides functionality to deal with multiple astronomical data sets.
    \begin{itemize}
      \item \texttt{astroNN.datasets} $-$ Besides multiple astronomical data sets like APOKASC (APOGEE-Kepler asteroseismology catalog), this contains ``Galaxy10'' which is an alternative to the standard MNIST or Cifar10 NN example training sets that is designed by us using data from Galaxy Zoo \citep{2008MNRAS.389.1179L} and SDSS \citep{2011ApJS..193...29A}. The data set consists of colored galaxy images and their morphology and it can be used to introduce astronomy researchers to deep-learning tools. The data set is available at \url{http://astro.utoronto.ca/~bovy/Galaxy10/Galaxy10.h5} with documentation at \url{https://astronn.readthedocs.io/en/v1.0.0/galaxy10.html}
	\end{itemize}
\end{itemize}

\subsection{Example of using Neural Net to infer parameters and abundances on arbitrary APOGEE spectra}

Besides providing general tools for deep learning in astronomy, we also share the actual networks trained and discussed in this paper in a separate \texttt{GitHub} repository associated with this paper. Here we give an example of how to use the main \verb|ApogeeBCNNCensored()| network for determining stellar parameters and abundances for a given APOGEE spectrum. First follow the following instructions:

\begin{enumerate}
  \item Install \texttt{astroNN} by following instructions from \url{https://astronn.readthedocs.io/en/v1.0.0/quick_start.html}
  \item Obtain the repository containing the code to reproduce all figures in this paper at \url{https://github.com/henrysky/astroNN_spectra_paper_figures}
  \item Open a python terminal under the repository folder but outside the folder \texttt{astroNN\_0617\_run001}
\end{enumerate}

Then copy and paste the following code to do inference with the neural net in this paper on 2M19060637+4717296, which is the spectrum shown in Figure \ref{figure:norm_aspcap}

\begin{lstlisting}[language=Python, caption={Example of using Neural Net to infer parameters and abundances on APOGEE spectra}]
from astropy.io import fits
from astroNN.apogee import visit_spectra, apogee_continuum
from astroNN.models import load_folder

# the same spectrum as used in figure 5
opened_fits = fits.open(visit_spectra(dr=14, apogee='2M19060637+4717296'))
spectrum = opened_fits[1].data
spectrum_err = opened_fits[2].data
spectrum_bitmask = opened_fits[3].data

# using default continuum and bitmask values to continuum normalize
norm_spec, norm_spec_err = apogee_continuum(spectrum, spectrum_err, 
                                            bitmask=spectrum_bitmask, dr=14)

# load neural net
neuralnet = load_folder('astroNN_0617_run001')

# inference, if there are multiple visits, then you should use the globally 
# weighted combined spectra (i.e. the second row)
pred, pred_err = neuralnet.test(norm_spec)

print(neuralnet.targetname)  # output neurons representation
print(pred)  # prediction
print(pred_err['total'])  # prediction uncertainty
\end{lstlisting}

\subsection{Fast Monte Carlo Inference on GPU}\label{subsec:fastMC}

For a probabilistic (i.e. a neural network that has different outputs every time you do inference on the same data) \texttt{keras} or \texttt{tensorflow.keras} neural network model which has a single input, as opposed to some keras multi-input model (in such case you need to concatenate multiple inputs and unpack the inputs in the model), to be inferred and a single output array which is the concatenation of the prediction and the predictive variance, you can wrap the model using \texttt{astroNN}'s \texttt{FastMCInference()} to get a new model, which gives you the mean prediction, predictive uncertainty, and model uncertainty with great performance on GPU. An example of this is in the following pseudo-code snippet

\begin{lstlisting}[language=Python, caption={Fast Monte Carlo Inference on GPU}]
from astroNN.nn.layers import FastMCInference

# keras_model is your model with 1 output which is a concatenation of all 
# label predictions and predictive variance
keras_model = Model(....)

# fast_mc_model is the new keras model capable of fast Monte-Carlo integration on GPU
# n=number of Monte-Carlo run
fast_mc_model = FastMCInference(keras_model, n=100)

# You can just use keras API with the new model such as
result = fast_mc_model.predict(.....)

# here is the result dimension
predictions = result[:, :(result.shape[1] // 2), 0]  # mean prediction
# model uncertainty
mc_dropout_uncertainty = result[:, :(result.shape[1] // 2), 1]
# predictive uncertainty
predictions_var = result[:, (result.shape[1] // 2):, 0]
\end{lstlisting}


 \bsp	
\label{lastpage}
\end{document}